\documentclass[aps, twocolumn, nofootinbib
]{revtex4-2}
\usepackage{pkgs}

\begin{document}
\title{Metastable Neutron Stars as Transient-Rate Detectors of Heavy Dark Matter}
\author{Ziwen Yin}
\email{ziwenyin@sjtu.edu.cn}
\author{Hong-Yi Zhang}
\email{hongyi18@sjtu.edu.cn}
\thanks{corresponding author}
\affiliation{Tsung-Dao Lee Institute \& School of Physics and Astronomy, Shanghai Jiao Tong University, Shanghai 201210, China}
\date{\today}

\begin{abstract}
Metastable neutron stars can undergo catastrophic conversion to quark-containing stars when a localized energy deposition nucleates a critical bubble. We show that dark-matter-induced conversions can be constrained by the present transient rate of the neutron star population, rather than by the survival of any individual hadronic star as often assumed in previous literature. Population depletion makes the present transition rate nonmonotonic: weak triggering produces few transitions, whereas sufficiently rapid triggering predominantly converts stars formed in the past. Consequently, transient observations can exclude a finite interval of interaction rates, qualitatively different from conventional one-sided survival bounds. Using the beaming-corrected short gamma ray burst rate as a benchmark, we derive constraints on dark matter annihilation, decay, inelastic particle scattering, and elastic scattering of macroscopic dark matter, including parameter space not covered by existing probes. Transient demographics thus provide a new population level probe of dark sector interactions and the phase structure of ultradense matter.
\end{abstract}
\maketitle

\begingroup
\renewcommand{\thefootnote}{}
\footnotetext{Both authors contributed equally to this work.}
\endgroup

\emph{Introduction---}Neutron stars (NSs) probe the phase structure of strongly interacting matter at baryon densities inaccessible to terrestrial experiments. At sufficiently high density, QCD predicts a transition from hadronic matter to deconfined quark matter, although whether this regime is reached in NS cores remains an open question \cite{Baym:2017whm, Annala:2019puf}. One viable possibility is a strong first-order hadron-quark transition \cite{Hsu:1998eu, Fodor:2004nz, Alford:2013aca}. Such a transition need not occur as soon as the coexistence pressure is crossed---surface energy can delay the formation of a critical droplet and leave a hadronic star metastable \cite{Iida:1998pi, Bombaci:2016xuj}. In nucleation models, the conversion time depends exponentially on the central pressure. It diverges at the equilibrium transition mass $M_0$ and can become shorter than the age of the universe above a model-dependent critical mass $M_\mrm{cr}$. Stars in the intervening window, $M_0<M_\star<M_\mrm{cr}$, may therefore survive for astrophysical times while remaining susceptible to a localized perturbation \cite{Bombaci:2016xuj}. Once a critical droplet forms, conversion to a hybrid or strange quark star can release energy of order $10^{53}\,\mathrm{erg}$ and may produce a short gamma ray burst (SGRB), a neutrino burst, and gravitational waves \cite{Perez-Garcia:2010xlt, Perez-Garcia:2013dwa, Drago:2015qwa, Bombaci:2016xuj, Marquez:2017bqm, Blas:2022xco}.

The extreme density, magnetic fields, escape velocity, and age of NSs also make them powerful detectors of dark matter (DM). Existing searches exploit DM capture, heating, cooling, or catastrophic accumulation to constrain DM interactions \cite{Baryakhtar:2017dbj, Bramante:2013nma, Zhang:2023vva, Raffelt:1996wa, Walters:2024vaw, Edwards:2020afl, Zhang:2023ktk,Yin:2026rfe}. A complementary possibility is that an energetic DM process deposits enough energy within a sufficiently small region to overcome the quark droplet nucleation barrier. DM annihilation inside NSs was first proposed to seed strange quark matter \cite{Perez-Garcia:2010xlt} and was later connected to a distinct population of SGRBs \cite{Perez-Garcia:2013dwa, Drago:2015qwa}. Subsequent work studied how annihilation energy modifies the tunneling rate \cite{Herrero:2019esf} and considered DM decay, annihilation, and scattering as possible triggers \cite{AngelesPerez-Garcia:2014cho,Bhutani:2025jfo}. These studies suggest that phase transitions in NSs can probe hidden sector processes inaccessible to terrestrial experiments, provided that the deposited energy is sufficiently localized.

Turning this sensitivity into a reliable constraint requires care about what observations actually establish. Previous analyses have assumed that a selected old compact star remains hadronic \cite{AngelesPerez-Garcia:2014cho, Bhutani:2025jfo}, or that no hadronic star in a chosen population has converted \cite{Bhutani:2025jfo}. Neither premise has yet been observationally established. Moreover, sufficiently massive hadronic stars may convert through spontaneous nucleation without any DM trigger \cite{Bombaci:2016xuj}, so single-star survival bounds require identifying both the metastable mass window and an observed hadronic star within it.

Here we instead constrain DM-induced conversions using the present transient rate of the metastable NS population. Because rapid triggering depletes this population, the present rate is nonmonotonic in the microscopic interaction rate, producing a two-sided excluded interval. This approach does not require any observed NS to be identified as hadronic. Using the observationally inferred SGRB rate \cite{Mandhai:2018cdl, Dichiara:2019kuw, Escorial:2022nvp, Fong:2015oha}, we apply this framework to particle and macroscopic DM, obtaining constraints not covered by current terrestrial and astrophysical probes.

\emph{Transition rate constraint from metastable neutron stars---}Consider old NSs in the metastable mass window $M_0<M_\star<M_\mrm{cr}$. We assume that each star has remained metastable for a characteristic time $t_\mrm{meta}$ and that DM-induced transitions form a Poisson process with a time-independent rate. For a normalized NS mass distribution $p_\mrm{NS}(M_\star)$, the present transition rate density is
\begin{align} 
\label{Rtr_general} 
\mcal R_\mrm{tr} = n_\mrm{NS}\int_{M_0}^{M_\mrm{cr}}dM_\star\, p_\mrm{NS}(M_\star)\lambda_\chi(M_\star) e^{-\lambda_\chi(M_\star)t_\mrm{meta}} ~,
\end{align} 
where $n_\mrm{NS}$ is the present-day comoving number density of old NSs and $\lambda_\chi$ is the DM-induced transition rate per star. Here, we neglect star formation since it has a small impact, see Sec.~\ref{sec:formation} of the Supplemental Material. Typically, $t_\mrm{meta} \lesssim t_\mrm{NS}$ since a NS may enter the metastable regime either at formation or only after subsequent evolution, such as spin down or accretion. The exponential factor is the probability that a star has not already converted. The transition rate can be written in terms of the differential rate of local energy deposition as \begin{align} 
\label{lambda_general} 
\lambda_\chi(M_\star) = \int dE_\mrm{dep}\, \frac{d\Gamma_\chi}{dE_\mrm{dep}}\, P_\mrm{bub}(M_\star,E_\mrm{dep}) ~,
\end{align} where $P_\mrm{bub}$ is the probability that an energy deposition $E_\mrm{dep}$ nucleates a supercritical quark matter bubble.

We first consider interactions for which every event counted in $\Gamma_\chi$ deposits enough energy within the nucleation region to trigger a transition, which will be discussed in more detail below. In this regime, $P_\mrm{bub}\simeq1$ and $\lambda_\chi\simeq\Gamma_\chi$. If $\Gamma_\chi$ varies negligibly across the metastable mass window, Eq.~\eqref{Rtr_general} reduces to 
\begin{align} 
\label{Rtr_benchmark} 
\mcal R_\mrm{tr} =n_\mrm{NS}f_\mrm{meta}\Gamma_\chi e^{-\Gamma_\chi t_\mrm{meta}} ~,
\end{align} 
where $f_\mrm{meta}\equiv\int_{M_0}^{M_\mrm{cr}} dM_\star\,p_\mrm{NS}(M_\star)$ is the fraction of NSs in the metastable window. For $\Gamma_\chi t_\mrm{meta}\ll1$, the rate grows linearly with $\Gamma_\chi$. It reaches a maximum at $\Gamma_\chi=t_\mrm{meta}^{-1}$ and then decreases because stars with larger transition rates were depleted earlier. Thus, both very slow triggering and rapid depletion can yield a small present-day rate.

Under the benchmark assumption that each conversion produces an observable transient, we require the transition rate to satisfy $\mcal R_\mrm{tr}< \mcal R_\mrm{SGRB}^\mrm{lim}$ with $\mcal R_\mrm{SGRB}^\mrm{lim} \equiv f_\mrm{iso} \mcal R_\mrm{SGRB}^\mrm{tot}$, where $\mcal R_\mrm{SGRB}^\mrm{tot}$ denotes the beaming-corrected total rate of SGRB-like transients and $f_\mrm{iso}$ is the fraction associated with unknown isolated sources. For simplicity, we conservatively set $f_\mrm{iso}=1$ throughout this work, while noting that future population studies may favor smaller values. Direct searches for nearby SGRBs constrain the detectable local rate \cite{Mandhai:2018cdl, Dichiara:2019kuw}, whereas afterglow jet angle measurements give beaming-corrected intrinsic rates \cite{Fong:2015oha, Escorial:2022nvp}. We adopt the upper end of these estimates, $\mcal R_\mrm{SGRB}^\mrm{tot} = 2\times10^3\, \mrm{Gpc}^{-3}\yr^{-1}$, as our benchmark \cite{Fong:2015oha, Escorial:2022nvp}. Cosmic remnant inventories give $n_\mrm{NS} \sim (2.5$--$5) \times 10^{15}\, \mrm{Gpc}^{-3}$ \cite{Fukugita:2004ee, Elbert:2017sbr, 2025MNRAS.540.2359M}; we take $n_\mrm{NS} = 3\times10^{15}\,\mrm{Gpc}^{-3}$.\footnote{Using the Milky-Way-equivalent galaxy density $n_\mrm{MWEG} = 0.0116 \, \Mpc^{-3}$ \cite{LIGOScientific:2010nhs}, this corresponds to $3\times 10^8$ NSs in the Milky Way, consistent with the galactic estimate $10^8$--$10^9$ \cite{2009PASP..121..814O, 2010A&A...510A..23S}.} Although $M_0$ and $M_\mrm{cr}$ depend on the equations of state, nucleation calculations typically find metastable windows of order $0.1M_\odot$ \cite{Bombaci:2016xuj}. Together with the inferred NS birth mass distribution \cite{You:2024bmk}, this motivates the benchmark $f_\mrm{meta}=0.01$. Given $t_\mrm{meta}\lesssim t_\mrm{NS}$, we take $t_\mrm{meta} \simeq 1\,\mrm{Gyr}$ as a representative value, leaving a population-level treatment of NS metastable times to future work. Putting everything together, the rate bound gives
\begin{align}
\label{Gamma_chi_constraint}
\Gamma_\chi<7\times10^{-11}\,\yr^{-1}
\quad\text{or}\quad
\Gamma_\chi>4\times10^{-9}\,\yr^{-1} ~.
\end{align}
The intermediate interval is excluded under our benchmark assumptions; see Sec.~\ref{sec:rate_boundary} of the Supplemental Material for more details. Fig.~\ref{fig:transitionrate} shows how the transition rate depends on the event rate per star, $\Gamma_\chi$, and on the population parameters. The benchmark choice (solid lines) excludes an intermediate range of $\Gamma_\chi$, while both sufficiently slow and sufficiently rapid triggering remain allowed. The upper allowed branch has no analogue in a single-star survival bound: when triggering is sufficiently rapid, nearly all metastable stars converted long before the present epoch. Moreover, a smaller metastable fraction $f_\mrm{meta}$ weakens the constraint, whereas improved identification of SGRB sources, hence a more precise determination of $f_\mrm{iso}$, could strengthen it in the future. Our method does not require any particular observed NS to be identified as hadronic. It instead relies on the existence of a metastable population and on its conversions producing SGRB-like transients. Next, we determine when a DM interaction deposits enough localized energy
for the approximation $\lambda_\chi\simeq\Gamma_\chi$ to hold.

\begin{figure}
\centering
\includegraphics[width=\linewidth]{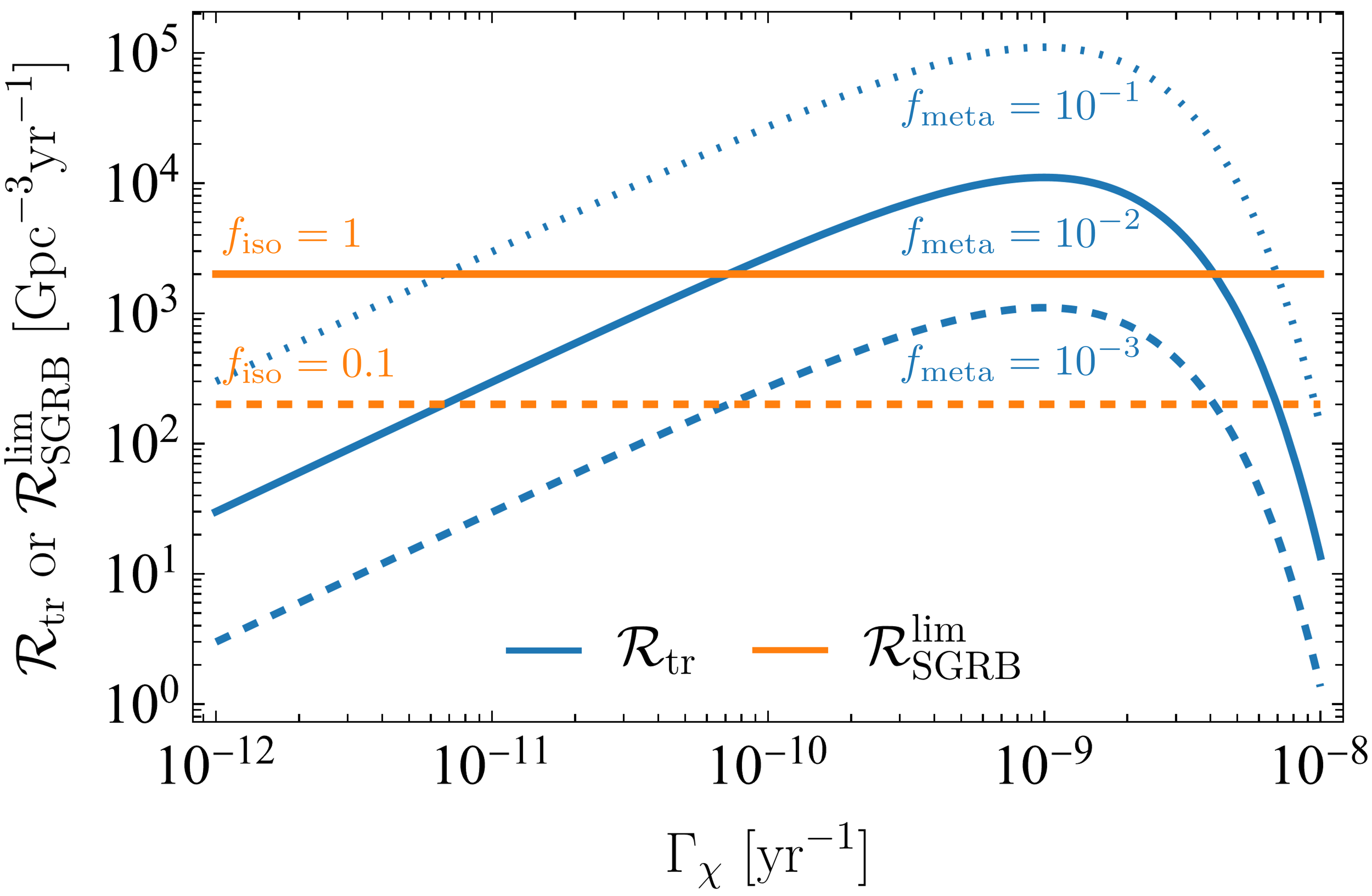}
\caption{Transition rate density as a function of the event rate per star, $\Gamma_\chi$. Blue curves show the predicted transition rate from Eq.~\eqref{Rtr_benchmark}, while orange curves show the observational upper limit from SGRB-like transients. Solid curves correspond to the benchmark parameters adopted in this work; dashed and dotted curves illustrate variations in the metastable NS fraction, $f_\mrm{meta}$, and the fraction of SGRBs associated with unknown isolated sources, $f_\mrm{iso}$. The benchmark choice excludes an intermediate range of $\Gamma_\chi$, while both sufficiently slow and sufficiently rapid triggering remain allowed.}
\label{fig:transitionrate}
\end{figure}

\emph{Dark-matter-induced bubble nucleation---}We now specify when a single DM interaction can trigger the transition. In the thin-wall Lifshitz-Kagan description, the potential energy of a quark matter droplet of radius $R$ is \cite{Iida:1998pi,Bombaci:2016xuj}
\begin{align}
U(R) = 4\pi\sigma_\mrm{S}R^2-\frac{4\pi}{3}\Delta\omega R^3 ~,
\end{align}
where $\sigma_\mrm{S}$ is the hadron-quark surface tension and $\Delta\omega=n_\mrm{Q}(\mu_\mrm{H}-\mu_\mrm{Q})>0$ is the bulk free energy gain per unit volume. Here, $n_\mrm{Q}$ is the baryon density of the quark phase, while $\mu_\mrm{H}$ and $\mu_\mrm{Q}$ are the baryon chemical potentials of the hadronic and quark phases evaluated at the same pressure and temperature. The potential reaches its maximum at the critical radius $R_\mrm{c}=2\sigma_\mrm{S}/\Delta\omega$, with barrier height $U_\mrm{max} = 16\pi \sigma_\mrm{S}^3/(3\Delta\omega^2)$,
\begin{align}
\label{bubble_barrier}
U_\mrm{max}
=4.5\,\GeV
\left(\frac{\sigma_\mrm{S}}{30\,\MeV/\mrm{fm}^2}\right)^3
\left(\frac{10\,\MeV/\mrm{fm}^3}{\Delta\omega}\right)^2 ~.
\end{align}
The surface tension is strongly model dependent; NS nucleation studies commonly use $\sigma_\mrm{S} = 10$--$30\,\MeV/\mrm{fm}^2$, with $30\,\MeV/\mrm{fm}^2$ serving as a conventional high-end benchmark \cite{Iida:1998pi, Bombaci:2016xuj, Herrero:2019esf}. Calculations for metastable matter yield barriers on the GeV scale, particularly when hyperons are present \cite{Iida:1998pi}, while recent hidden sector studies have considered barriers of order $10\,\GeV$ \cite{Bhutani:2025jfo}. Therefore, we adopt the illustrative threshold $U_\mrm{max}=10\,\GeV$. The critical radius is
\begin{align}
\label{critical_radius}
R_\mrm{c}=8.9\,\mrm{fm}
\left(\frac{U_\mrm{max}}{10\,\GeV}\right)^{1/2}
\left(\frac{30\,\MeV/\mrm{fm}^2}{\sigma_\mrm{S}}\right)^{1/2} ~.
\end{align}
An energy deposition that raises a fluctuation above the barrier produces a supercritical droplet of the size $R\sim R_\mrm{c}$, which subsequently grows classically. Changing this benchmark shifts the minimum deposited energy, and hence the minimum DM mass, to which our constraints apply, but does not alter the population rate argument.

The relevant quantity is the energy $E_\mrm{dep}$ deposited within a region of size $\lesssim R_\mrm{c}$, rather than the total energy released by the interaction. For hadronically interacting final states, a useful estimate of the deposition length is the inelastic mean free path, $l_\mrm{dep}^\mrm{had} \sim (n_\mrm{b}\sigma_\mrm{inel})^{-1} \simeq 0.7\,\mrm{fm}
(0.15\,\mrm{fm}^{-3}/n_\mrm{b}) ( 100\,\mrm{mb} / \sigma_\mrm{inel} )$, where $n_\mrm{b}$ is the local baryon density and $\sigma_\mrm{inel}$ is an effective inelastic hadron-nucleon cross section. Since $l_\mrm{dep}^\mrm{had} < R_\mrm{c}$ for the benchmark parameters, we treat the hadronic component of the injected energy as locally deposited. Within this localized threshold approximation, the bubble nucleation probability is approximated as $P_\mrm{bub}(E_\mrm{dep}) \simeq \theta( E_\mrm{dep} - E_\mrm{th} )$, where $\theta$ is the Heaviside step function and we assume the threshold energy $E_\mrm{th}\sim U_\mrm{max}$ following \cite{Herrero:2019esf, AngelesPerez-Garcia:2014cho, Bhutani:2025jfo}. Thus, if every event included in $\Gamma_\chi$ deposits $E_\mrm{dep}>U_\mrm{max}$ within the critical region, then $\lambda_\chi\simeq\Gamma_\chi$, as assumed above. We comment that a larger threshold energy would shift the the low-mass cutoff of below constraints for DM annihilation, decay, and inelastic scattering. The microscopic nucleation calculation and the limitations of the localized deposition approximation are discussed in Sec.~\ref{sec:nucleation} of the Supplemental Material.

\emph{Constraints on particle and macroscopic dark matter---}The population-level constraint \eqref{Gamma_chi_constraint} applies to any process once its per-star triggering rate is specified. We now translate it into constraints on representative DM processes. For an asymptotic Maxwellian distribution with velocity dispersion $v_0$, gravitational focusing enhances the unbound DM density in the NS core to \cite{Bromley:2011aa}
\begin{align}
n_{\chi,\mrm{core}}\simeq 9\times10^2
\frac{\rho_{\chi,\infty}}{m_\chi}
\left(\frac{v_\mrm{esc}}{0.6c}\right)
\left(\frac{220\,\km/\s}{v_0}\right) ~,
\end{align}
where $m_\chi$ and $\rho_{\chi,\infty}$ are the DM mass and asymptotic density, $v_\mrm{esc}$ is the escape velocity of NSs, and $v_0$ is the asymptotic velocity dispersion of DM. As a representative halo environment, we take $\rho_{\chi,\infty} = 0.4\,\GeV/\cm^3$ and $v_0=220\,\km/\s$, together with the canonical NS value $v_\mrm{esc}=0.6c$. We approximate the metastable central region by a volume $V_\mrm{meta}\simeq 10^6\,\m^3$, where the density and pressure may be regarded as constant \cite{Iida:1998pi}.\footnote{In \cite{Bhutani:2025jfo}, the authors identified the relevant metastable region as a much larger volume within which quark matter is energetically favored. In this case, however, the potential barrier at the boundaries is considerably higher than the core value $U_\mrm{max}$.} We further assume that the energy above threshold is carried by hadronically interacting particles and deposited within the critical radius. Details of the focusing and subsequent interaction rate calculations are given in Sec.~\ref{sec:dm_constraints} and \ref{sec:macro} of the Supplemental Material.

For $s$-wave annihilation into the Standard Model (SM) particles, $\chi\chi\to\mrm{SM}$, the threshold condition is $2m_\chi\gtrsim U_\mrm{max}$, and the event rate per star is $ \Gamma_\chi^\mrm{ann} \simeq n_{\chi,\mrm{core}}^2 \langle\sigma v\rangle_\mrm{ann}V_\mrm{meta} / 2$. Here $\langle\sigma v\rangle_\mrm{ann}$ is the velocity-averaged annihilation cross section and the factor $1/2$ avoids double counting identical initial-state particles. Eq.~\eqref{Gamma_chi_constraint} then excludes
\begin{align}
8 \times10^{-34} < \frac{\langle\sigma v\rangle_\mrm{ann}}{\cm^3/\s} \left(\frac{5\,\GeV}{m_\chi}\right)^2 <5 \times10^{-32} ~,
\label{ann_constraint}
\end{align}
for $m_\chi\gtrsim5\,\GeV$ at our benchmark $U_\mrm{max}=10\,\GeV$. 
\begin{figure}
\centering
\includegraphics[width=\linewidth]{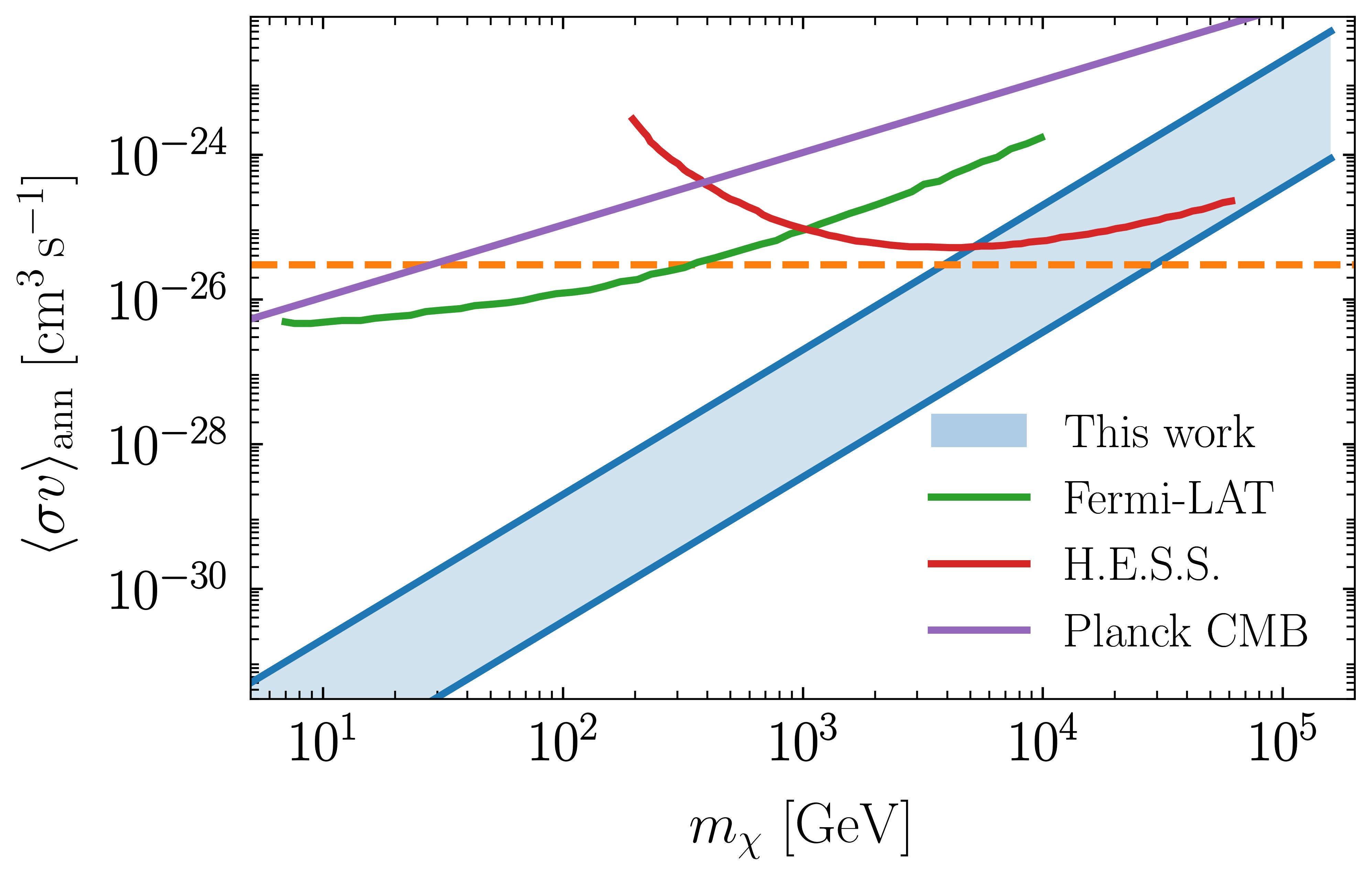}
\caption{Constraints on $s$-wave DM annihilation into $b\bar b$. The blue band indicates the NS phase transition constraint under the benchmark assumptions described in the text, which also applies to other hadronic final states. The other colored curves show representative limits from Fermi-LAT observations of dwarf spheroidal galaxies, the H.E.S.S. Inner Galaxy survey, and Planck \cite{McDaniel:2023bju, HESS:2022ygk, Planck:2018vyg}. The orange dashed line denotes the canonical thermal relic annihilation cross section $\la \sigma v\ra = 3\times 10^{-26} \cm^3 \s^{-1}$ \cite{Steigman:2012nb}.}
\label{fig:annihilationconstraint}
\end{figure}
In Fig.~\ref{fig:annihilationconstraint}, we show the constraint from Eq.~\eqref{ann_constraint} together with existing bounds on velocity-independent DM annihilation into $b\bar b$. Our constraint probes a region of parameter space complementary to conventional indirect searches. For comparison, we also show representative limits from gamma ray observations of dwarf spheroidal galaxies, the Galactic center, the cosmic microwave background, as well as the canonical thermal relic annihilation cross section.\footnote{More precise calculations yield a mass-dependent thermal relic annihilation cross section \cite{Steigman:2012nb, Baum:2016oow}.} Notably, our constraint disfavors canonical thermal relic DM in the mass range $4\times 10^3 \,\GeV \lesssim m_\chi \lesssim 3\times 10^4 \,\GeV$.

\begin{figure}
\centering
\includegraphics[width=\linewidth]{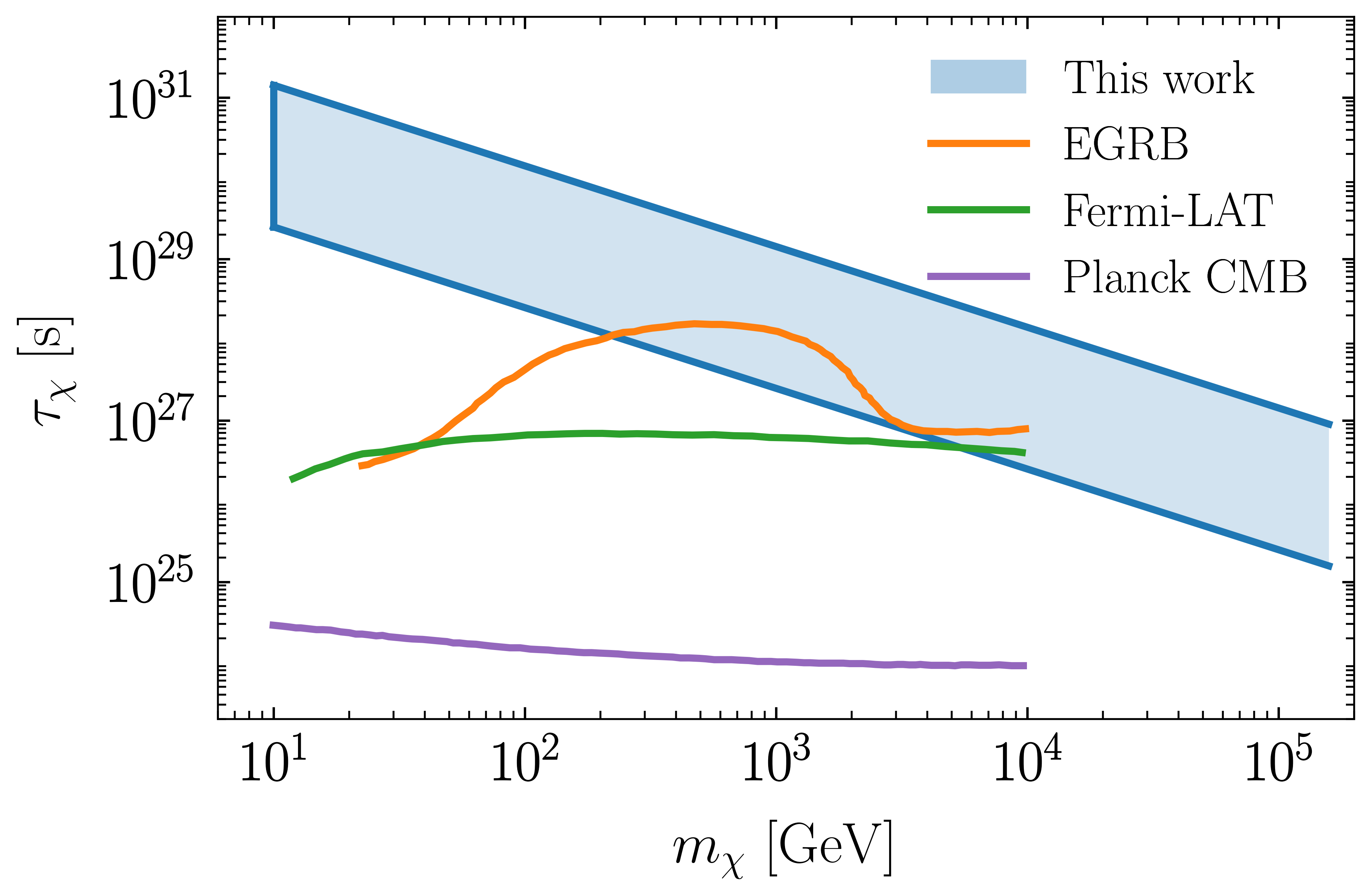}
\caption{Constraints on the lifetime of DM decaying into $b\bar b$. The blue band shows the region excluded by the NS phase transition criterion, which also applies to other hadronic final states. Solid curves show representative constraints from the extragalactic gamma ray background (EGRB), Fermi-LAT observations of dwarf spheroidal galaxies, and Planck \cite{Ando:2015qda, Baring:2015sza, Slatyer:2016qyl}.}
\label{fig:decayconstraint}
\end{figure}

For decaying DM, $\chi\to\mrm{SM}$, the event rate is
$\Gamma_\chi^\mrm{dec}\simeq n_{\chi,\mrm{core}}V_\mrm{meta}/\tau_\chi$, where $\tau_\chi$ is the DM lifetime. For $m_\chi\gtrsim U_\mrm{max}$, the excluded lifetime interval is
\begin{align}
3 \times10^{29} < \frac{\tau_\chi}{\s} \left(\frac{m_\chi}{10\,\GeV}\right) <2\times10^{31} ~.
\label{dec_constraint}
\end{align}
At longer lifetimes, decays are too rare to overproduce present-day transitions; at shorter lifetimes, the metastable population was depleted at earlier times. In Fig.~\ref{fig:decayconstraint}, we show the constraint on the DM lifetime from NS phase transitions together with representative existing bounds. The NS constraint excludes a broad range of lifetimes, with sensitivity complementary to conventional indirect searches based on gamma ray and cosmic microwave background observations.

An inelastic process $\chi + N\to\chi' + \mrm{SM}$, where $N$ denotes a nucleon, can trigger a transition if the visible final state carries sufficient energy. Up to corrections from the incident kinetic energy and the nucleon mass, this requires $m_\chi-m_{\chi'}\gtrsim U_\mrm{max}$. For a baryon number density $n_\mrm{b}=0.15\,\mrm{fm}^{-3}$, the event rate is
$\Gamma_\chi^\mrm{inel}\simeq
n_\mrm{b}n_{\chi,\mrm{core}}\sigma_\mrm{inel}
v_\mrm{esc}V_\mrm{meta}$, where $\sigma_\mrm{inel}$ is the inelastic DM-nucleon cross section evaluated at the infall velocity $v_\mrm{esc}$. Eq.~\eqref{Gamma_chi_constraint} then excludes
\begin{align}
2 \times 10^{-80} < \frac{\sigma_\mrm{inel}}{\cm^2}
\left(\frac{10\,\GeV}{m_\chi}\right) <1 \times10^{-78} ~.
\label{inel_constraint}
\end{align}
The strong sensitivity arises from the large number of nucleons in the metastable core.

Macroscopic DM can instead deposit the threshold energy through several elastic scatters within a region of size $R_\mrm{c}$. In the optically thick macro limit, the effective macro-nucleon cross section approaches its geometric value, $\sigma_\mrm{macro}=\pi R_\chi^2$ \cite{Jacobs:2014yca, SinghSidhu:2019tbr, Bramante:2026wzh}. The rate at which macros trigger transitions is $\Gamma_\chi^\mrm{macro} \simeq n_{\chi,\mrm{core}} \pi R_\mrm{meta}^2 v_\mrm{esc} P_\mrm{th}$, where $R_\mrm{meta} \sim V_\mrm{meta}^{1/3}$ and $P_\mrm{th}$ is the probability that a transiting macro deposits at least $U_\mrm{max}$ within one of the approximately $R_\mrm{meta}/R_\mrm{c}$ critical-size domains along its path. Because $m_\chi\gg m_N$, the macro trajectory is approximately unchanged by individual scatters. The mean number of scatters within one domain is $\lambda_\mrm{c} = n_\mrm{b} \sigma_\mrm{macro} R_\mrm{c}$. Each scatter transfers a characteristic energy $E_\mrm{recoil}\sim m_N(\gamma_\mrm{esc}^2-1)$, where $\gamma_\mrm{esc} = (1-v_\mrm{esc}^2)^{-1/2}$. Then the number of scatters required to cross the nucleation barrier is $N_\mrm{th}\simeq U_\mrm{max}/E_\mrm{recoil}\simeq 19$ for the benchmark parameters. In the small-$\lambda_\mrm{c}$ tail of the Poisson distribution, $P_\mrm{th}\simeq(R_\mrm{meta} / R_\mrm{c}) \lambda_\mrm{c}^{N_\mrm{th}} / N_\mrm{th}!$. Consequently, $\Gamma_\chi^\mrm{macro} \propto (\rho_{\chi,\infty}/m_\chi)\sigma_\mrm{macro}^{19}$ and hence $\sigma_\mrm{macro} \propto (m_\chi\Gamma_\chi^\mrm{macro})^{1/19}$. Applying Eq.~\eqref{Gamma_chi_constraint} then excludes
\begin{align}
1.3\times10^{-27} \lesssim \frac{\sigma_\mrm{macro}}{\cm^2} 
\left(\frac{1\,\g}{m_\chi}\right)^{1/19} \lesssim
1.6\times10^{-27}  ~,
\label{macro_constraint}
\end{align}
for the benchmark parameters specified above; the full parameter dependence is given in Sec.~\ref{sec:macro} of the Supplemental Material. Eq.~\eqref{macro_constraint} assumes that the macro reaches the core without substantial deceleration. This requires
$\sigma_\mrm{macro} / m_\chi \lesssim 4\times10^{-21}\,\cm^2/\g$ and hence $m_\chi \gtrsim 4\times10^{-7}\,\g$ within the excluded band. Our elastic scattering treatment also requires the macro to be larger than its Schwarzschild radius. Since $\sigma_\mrm{macro}=\pi R_\chi^2$, this condition is $\sigma_\mrm{macro}/m_\chi^2 > 7\times 10^{-56} \cm^2/\g^2$, which restricts Eq.~\eqref{macro_constraint} to $m_\chi \lesssim 2 \times10^{14}\,\g$.

\emph{Conclusion---}We have shown that metastable NSs can probe dark sector interactions through the present rate of catastrophic phase transitions, rather than through the survival of any particular hadronic star \cite{AngelesPerez-Garcia:2014cho, Bhutani:2025jfo}. For a population of characteristic metastable age $t_\mrm{meta}$, depletion changes the rate from the naive scaling $\mcal R_\mrm{tr}\propto\Gamma_\chi$ to $\mcal R_\mrm{tr}\propto\Gamma_\chi e^{-\Gamma_\chi t_\mrm{meta}}$, making the present transition rate nonmonotonic in the microscopic triggering rate. Consequently, transient observations can exclude a two-sided interval of $\Gamma_\chi$: weak triggering produces too few events, while sufficiently rapid triggering depletes the metastable population at earlier times. This upper allowed branch has no analogue in conventional single-star survival bounds.

Using the beaming-corrected SGRB rate as a conservative ceiling, we translated this population constraint into two-sided excluded regions for DM annihilation, decay, inelastic scattering, and elastic scattering of macroscopic DM. The numerical reach depends on the abundance and age distribution of metastable stars, the localized nucleation threshold, and the observable transient accompanying conversion, but the underlying rate and depletion mechanism does not. Improved modeling of these ingredients, together with gamma ray, neutrino, and gravitational wave searches for conversion transients \cite{Perez-Garcia:2013dwa, Drago:2015qwa, Marquez:2017bqm}, can sharpen the corresponding constraints.

More broadly, the same reasoning applies whenever rare interactions irreversibly transform a longlived astrophysical population. Transient demographics can therefore probe new physics even when the progenitor objects cannot be identified individually.

\begin{acknowledgments}
We would like to thank Wenyuan Ai and Luca Visinelli for insightful comments on the manuscript, and Zhiqiang Miao for helpful discussions. We acknowledge support from the Shanghai Magnolia Plan Pujiang Program 25PJA073.
\end{acknowledgments}

\bibliography{ref}

\begin{thebibliography}{53}%
\makeatletter
\providecommand \@ifxundefined [1]{%
 \@ifx{#1\undefined}
}%
\providecommand \@ifnum [1]{%
 \ifnum #1\expandafter \@firstoftwo
 \else \expandafter \@secondoftwo
 \fi
}%
\providecommand \@ifx [1]{%
 \ifx #1\expandafter \@firstoftwo
 \else \expandafter \@secondoftwo
 \fi
}%
\providecommand \natexlab [1]{#1}%
\providecommand \enquote  [1]{``#1''}%
\providecommand \bibnamefont  [1]{#1}%
\providecommand \bibfnamefont [1]{#1}%
\providecommand \citenamefont [1]{#1}%
\providecommand \href@noop [0]{\@secondoftwo}%
\providecommand \href [0]{\begingroup \@sanitize@url \@href}%
\providecommand \@href[1]{\@@startlink{#1}\@@href}%
\providecommand \@@href[1]{\endgroup#1\@@endlink}%
\providecommand \@sanitize@url [0]{\catcode `\\12\catcode `\$12\catcode
  `\&12\catcode `\#12\catcode `\^12\catcode `\_12\catcode `\%12\relax}%
\providecommand \@@startlink[1]{}%
\providecommand \@@endlink[0]{}%
\providecommand \url  [0]{\begingroup\@sanitize@url \@url }%
\providecommand \@url [1]{\endgroup\@href {#1}{\urlprefix }}%
\providecommand \urlprefix  [0]{URL }%
\providecommand \Eprint [0]{\href }%
\providecommand \doibase [0]{https://doi.org/}%
\providecommand \selectlanguage [0]{\@gobble}%
\providecommand \bibinfo  [0]{\@secondoftwo}%
\providecommand \bibfield  [0]{\@secondoftwo}%
\providecommand \translation [1]{[#1]}%
\providecommand \BibitemOpen [0]{}%
\providecommand \bibitemStop [0]{}%
\providecommand \bibitemNoStop [0]{.\EOS\space}%
\providecommand \EOS [0]{\spacefactor3000\relax}%
\providecommand \BibitemShut  [1]{\csname bibitem#1\endcsname}%
\let\auto@bib@innerbib\@empty
\bibitem [{\citenamefont {Baym}\ \emph {et~al.}(2018)\citenamefont {Baym},
  \citenamefont {Hatsuda}, \citenamefont {Kojo}, \citenamefont {Powell},
  \citenamefont {Song},\ and\ \citenamefont {Takatsuka}}]{Baym:2017whm}%
  \BibitemOpen
  \bibfield  {author} {\bibinfo {author} {\bibfnamefont {G.}~\bibnamefont
  {Baym}}, \bibinfo {author} {\bibfnamefont {T.}~\bibnamefont {Hatsuda}},
  \bibinfo {author} {\bibfnamefont {T.}~\bibnamefont {Kojo}}, \bibinfo {author}
  {\bibfnamefont {P.~D.}\ \bibnamefont {Powell}}, \bibinfo {author}
  {\bibfnamefont {Y.}~\bibnamefont {Song}},\ and\ \bibinfo {author}
  {\bibfnamefont {T.}~\bibnamefont {Takatsuka}},\ }\bibfield  {title} {\bibinfo
  {title} {{From hadrons to quarks in neutron stars: a review}},\ }\href
  {https://doi.org/10.1088/1361-6633/aaae14} {\bibfield  {journal} {\bibinfo
  {journal} {Rept. Prog. Phys.}\ }\textbf {\bibinfo {volume} {81}},\ \bibinfo
  {pages} {056902} (\bibinfo {year} {2018})},\ \Eprint
  {https://arxiv.org/abs/1707.04966} {arXiv:1707.04966 [astro-ph.HE]}
  \BibitemShut {NoStop}%
\bibitem [{\citenamefont {Annala}\ \emph {et~al.}(2020)\citenamefont {Annala},
  \citenamefont {Gorda}, \citenamefont {Kurkela}, \citenamefont
  {N{\"a}ttil{\"a}},\ and\ \citenamefont {Vuorinen}}]{Annala:2019puf}%
  \BibitemOpen
  \bibfield  {author} {\bibinfo {author} {\bibfnamefont {E.}~\bibnamefont
  {Annala}}, \bibinfo {author} {\bibfnamefont {T.}~\bibnamefont {Gorda}},
  \bibinfo {author} {\bibfnamefont {A.}~\bibnamefont {Kurkela}}, \bibinfo
  {author} {\bibfnamefont {J.}~\bibnamefont {N{\"a}ttil{\"a}}},\ and\ \bibinfo
  {author} {\bibfnamefont {A.}~\bibnamefont {Vuorinen}},\ }\bibfield  {title}
  {\bibinfo {title} {{Evidence for quark-matter cores in massive neutron
  stars}},\ }\href {https://doi.org/10.1038/s41567-020-0914-9} {\bibfield
  {journal} {\bibinfo  {journal} {Nature Phys.}\ }\textbf {\bibinfo {volume}
  {16}},\ \bibinfo {pages} {907} (\bibinfo {year} {2020})},\ \Eprint
  {https://arxiv.org/abs/1903.09121} {arXiv:1903.09121 [astro-ph.HE]}
  \BibitemShut {NoStop}%
\bibitem [{\citenamefont {Hsu}\ and\ \citenamefont
  {Schwetz}(1998)}]{Hsu:1998eu}%
  \BibitemOpen
  \bibfield  {author} {\bibinfo {author} {\bibfnamefont {S.~D.~H.}\
  \bibnamefont {Hsu}}\ and\ \bibinfo {author} {\bibfnamefont {M.}~\bibnamefont
  {Schwetz}},\ }\bibfield  {title} {\bibinfo {title} {{On the QCD phase
  transition at finite baryon density}},\ }\href
  {https://doi.org/10.1016/S0370-2693(98)00615-7} {\bibfield  {journal}
  {\bibinfo  {journal} {Phys. Lett. B}\ }\textbf {\bibinfo {volume} {432}},\
  \bibinfo {pages} {203} (\bibinfo {year} {1998})},\ \Eprint
  {https://arxiv.org/abs/hep-ph/9803386} {arXiv:hep-ph/9803386} \BibitemShut
  {NoStop}%
\bibitem [{\citenamefont {Fodor}\ and\ \citenamefont
  {Katz}(2004)}]{Fodor:2004nz}%
  \BibitemOpen
  \bibfield  {author} {\bibinfo {author} {\bibfnamefont {Z.}~\bibnamefont
  {Fodor}}\ and\ \bibinfo {author} {\bibfnamefont {S.~D.}\ \bibnamefont
  {Katz}},\ }\bibfield  {title} {\bibinfo {title} {{Critical point of QCD at
  finite T and mu, lattice results for physical quark masses}},\ }\href
  {https://doi.org/10.1088/1126-6708/2004/04/050} {\bibfield  {journal}
  {\bibinfo  {journal} {JHEP}\ }\textbf {\bibinfo {volume} {04}},\ \bibinfo
  {pages} {050}},\ \Eprint {https://arxiv.org/abs/hep-lat/0402006}
  {arXiv:hep-lat/0402006} \BibitemShut {NoStop}%
\bibitem [{\citenamefont {Alford}\ \emph {et~al.}(2013)\citenamefont {Alford},
  \citenamefont {Han},\ and\ \citenamefont {Prakash}}]{Alford:2013aca}%
  \BibitemOpen
  \bibfield  {author} {\bibinfo {author} {\bibfnamefont {M.~G.}\ \bibnamefont
  {Alford}}, \bibinfo {author} {\bibfnamefont {S.}~\bibnamefont {Han}},\ and\
  \bibinfo {author} {\bibfnamefont {M.}~\bibnamefont {Prakash}},\ }\bibfield
  {title} {\bibinfo {title} {{Generic conditions for stable hybrid stars}},\
  }\href {https://doi.org/10.1103/PhysRevD.88.083013} {\bibfield  {journal}
  {\bibinfo  {journal} {Phys. Rev. D}\ }\textbf {\bibinfo {volume} {88}},\
  \bibinfo {pages} {083013} (\bibinfo {year} {2013})},\ \Eprint
  {https://arxiv.org/abs/1302.4732} {arXiv:1302.4732 [astro-ph.SR]}
  \BibitemShut {NoStop}%
\bibitem [{\citenamefont {Iida}\ and\ \citenamefont
  {Sato}(1998)}]{Iida:1998pi}%
  \BibitemOpen
  \bibfield  {author} {\bibinfo {author} {\bibfnamefont {K.}~\bibnamefont
  {Iida}}\ and\ \bibinfo {author} {\bibfnamefont {K.}~\bibnamefont {Sato}},\
  }\bibfield  {title} {\bibinfo {title} {{Effects of hyperons on the dynamical
  deconfinement transition in cold neutron star matter}},\ }\href
  {https://doi.org/10.1103/PhysRevC.58.2538} {\bibfield  {journal} {\bibinfo
  {journal} {Phys. Rev. C}\ }\textbf {\bibinfo {volume} {58}},\ \bibinfo
  {pages} {2538} (\bibinfo {year} {1998})},\ \Eprint
  {https://arxiv.org/abs/nucl-th/9808056} {arXiv:nucl-th/9808056} \BibitemShut
  {NoStop}%
\bibitem [{\citenamefont {Bombaci}\ \emph {et~al.}(2016)\citenamefont
  {Bombaci}, \citenamefont {Logoteta}, \citenamefont {Vida{\~n}a},\ and\
  \citenamefont {Provid{\^e}ncia}}]{Bombaci:2016xuj}%
  \BibitemOpen
  \bibfield  {author} {\bibinfo {author} {\bibfnamefont {I.}~\bibnamefont
  {Bombaci}}, \bibinfo {author} {\bibfnamefont {D.}~\bibnamefont {Logoteta}},
  \bibinfo {author} {\bibfnamefont {I.}~\bibnamefont {Vida{\~n}a}},\ and\
  \bibinfo {author} {\bibfnamefont {C.}~\bibnamefont {Provid{\^e}ncia}},\
  }\bibfield  {title} {\bibinfo {title} {{Quark matter nucleation in neutron
  stars and astrophysical implications}},\ }\href
  {https://doi.org/10.1140/epja/i2016-16058-5} {\bibfield  {journal} {\bibinfo
  {journal} {Eur. Phys. J. A}\ }\textbf {\bibinfo {volume} {52}},\ \bibinfo
  {pages} {58} (\bibinfo {year} {2016})},\ \Eprint
  {https://arxiv.org/abs/1601.04559} {arXiv:1601.04559 [astro-ph.HE]}
  \BibitemShut {NoStop}%
\bibitem [{\citenamefont {Perez-Garcia}\ \emph {et~al.}(2010)\citenamefont
  {Perez-Garcia}, \citenamefont {Silk},\ and\ \citenamefont
  {Stone}}]{Perez-Garcia:2010xlt}%
  \BibitemOpen
  \bibfield  {author} {\bibinfo {author} {\bibfnamefont {M.~A.}\ \bibnamefont
  {Perez-Garcia}}, \bibinfo {author} {\bibfnamefont {J.}~\bibnamefont {Silk}},\
  and\ \bibinfo {author} {\bibfnamefont {J.~R.}\ \bibnamefont {Stone}},\
  }\bibfield  {title} {\bibinfo {title} {{Dark matter, neutron stars and
  strange quark matter}},\ }\href
  {https://doi.org/10.1103/PhysRevLett.105.141101} {\bibfield  {journal}
  {\bibinfo  {journal} {Phys. Rev. Lett.}\ }\textbf {\bibinfo {volume} {105}},\
  \bibinfo {pages} {141101} (\bibinfo {year} {2010})},\ \Eprint
  {https://arxiv.org/abs/1007.1421} {arXiv:1007.1421 [astro-ph.CO]}
  \BibitemShut {NoStop}%
\bibitem [{\citenamefont {Perez-Garcia}\ \emph {et~al.}(2013)\citenamefont
  {Perez-Garcia}, \citenamefont {Daigne},\ and\ \citenamefont
  {Silk}}]{Perez-Garcia:2013dwa}%
  \BibitemOpen
  \bibfield  {author} {\bibinfo {author} {\bibfnamefont {M.~A.}\ \bibnamefont
  {Perez-Garcia}}, \bibinfo {author} {\bibfnamefont {F.}~\bibnamefont
  {Daigne}},\ and\ \bibinfo {author} {\bibfnamefont {J.}~\bibnamefont {Silk}},\
  }\bibfield  {title} {\bibinfo {title} {{Short GRBs and dark matter seeding in
  neutron stars}},\ }\href {https://doi.org/10.1088/0004-637X/768/2/145}
  {\bibfield  {journal} {\bibinfo  {journal} {Astrophys. J.}\ }\textbf
  {\bibinfo {volume} {768}},\ \bibinfo {pages} {145} (\bibinfo {year}
  {2013})},\ \Eprint {https://arxiv.org/abs/1303.2697} {arXiv:1303.2697
  [astro-ph.HE]} \BibitemShut {NoStop}%
\bibitem [{\citenamefont {Drago}\ \emph {et~al.}(2016)\citenamefont {Drago},
  \citenamefont {Lavagno}, \citenamefont {Metzger},\ and\ \citenamefont
  {Pagliara}}]{Drago:2015qwa}%
  \BibitemOpen
  \bibfield  {author} {\bibinfo {author} {\bibfnamefont {A.}~\bibnamefont
  {Drago}}, \bibinfo {author} {\bibfnamefont {A.}~\bibnamefont {Lavagno}},
  \bibinfo {author} {\bibfnamefont {B.}~\bibnamefont {Metzger}},\ and\ \bibinfo
  {author} {\bibfnamefont {G.}~\bibnamefont {Pagliara}},\ }\bibfield  {title}
  {\bibinfo {title} {{Quark deconfinement and the duration of short Gamma Ray
  Bursts}},\ }\href {https://doi.org/10.1103/PhysRevD.93.103001} {\bibfield
  {journal} {\bibinfo  {journal} {Phys. Rev. D}\ }\textbf {\bibinfo {volume}
  {93}},\ \bibinfo {pages} {103001} (\bibinfo {year} {2016})},\ \Eprint
  {https://arxiv.org/abs/1510.05581} {arXiv:1510.05581 [astro-ph.HE]}
  \BibitemShut {NoStop}%
\bibitem [{\citenamefont {Marquez}\ and\ \citenamefont
  {Menezes}(2017)}]{Marquez:2017bqm}%
  \BibitemOpen
  \bibfield  {author} {\bibinfo {author} {\bibfnamefont {K.~D.}\ \bibnamefont
  {Marquez}}\ and\ \bibinfo {author} {\bibfnamefont {D.~P.}\ \bibnamefont
  {Menezes}},\ }\bibfield  {title} {\bibinfo {title} {{Phase transition in
  compact stars: nucleation mechanism and $\gamma$-ray bursts revisited}},\
  }\href {https://doi.org/10.1088/1475-7516/2017/12/028} {\bibfield  {journal}
  {\bibinfo  {journal} {JCAP}\ }\textbf {\bibinfo {volume} {12}},\ \bibinfo
  {pages} {028}},\ \Eprint {https://arxiv.org/abs/1709.07040} {arXiv:1709.07040
  [astro-ph.HE]} \BibitemShut {NoStop}%
\bibitem [{\citenamefont {Blas}\ \emph {et~al.}(2026)\citenamefont {Blas},
  \citenamefont {Casalderrey-Solana}, \citenamefont {Mateos},\ and\
  \citenamefont {Sanchez-Garitaonandia}}]{Blas:2022xco}%
  \BibitemOpen
  \bibfield  {author} {\bibinfo {author} {\bibfnamefont {D.}~\bibnamefont
  {Blas}}, \bibinfo {author} {\bibfnamefont {J.}~\bibnamefont
  {Casalderrey-Solana}}, \bibinfo {author} {\bibfnamefont {D.}~\bibnamefont
  {Mateos}},\ and\ \bibinfo {author} {\bibfnamefont {M.}~\bibnamefont
  {Sanchez-Garitaonandia}},\ }\bibfield  {title} {\bibinfo {title} {{Megahertz
  Gravitational Waves from Neutron Star Mergers}},\ }\href
  {https://doi.org/10.1103/6yz9-94ql} {\bibfield  {journal} {\bibinfo
  {journal} {Phys. Rev. Lett.}\ }\textbf {\bibinfo {volume} {136}},\ \bibinfo
  {pages} {101401} (\bibinfo {year} {2026})},\ \Eprint
  {https://arxiv.org/abs/2210.03171} {arXiv:2210.03171 [hep-th]} \BibitemShut
  {NoStop}%
\bibitem [{\citenamefont {Baryakhtar}\ \emph {et~al.}(2017)\citenamefont
  {Baryakhtar}, \citenamefont {Bramante}, \citenamefont {Li}, \citenamefont
  {Linden},\ and\ \citenamefont {Raj}}]{Baryakhtar:2017dbj}%
  \BibitemOpen
  \bibfield  {author} {\bibinfo {author} {\bibfnamefont {M.}~\bibnamefont
  {Baryakhtar}}, \bibinfo {author} {\bibfnamefont {J.}~\bibnamefont
  {Bramante}}, \bibinfo {author} {\bibfnamefont {S.~W.}\ \bibnamefont {Li}},
  \bibinfo {author} {\bibfnamefont {T.}~\bibnamefont {Linden}},\ and\ \bibinfo
  {author} {\bibfnamefont {N.}~\bibnamefont {Raj}},\ }\bibfield  {title}
  {\bibinfo {title} {{Dark Kinetic Heating of Neutron Stars and An Infrared
  Window On WIMPs, SIMPs, and Pure Higgsinos}},\ }\href
  {https://doi.org/10.1103/PhysRevLett.119.131801} {\bibfield  {journal}
  {\bibinfo  {journal} {Phys. Rev. Lett.}\ }\textbf {\bibinfo {volume} {119}},\
  \bibinfo {pages} {131801} (\bibinfo {year} {2017})},\ \Eprint
  {https://arxiv.org/abs/1704.01577} {arXiv:1704.01577 [hep-ph]} \BibitemShut
  {NoStop}%
\bibitem [{\citenamefont {Bramante}\ \emph {et~al.}(2014)\citenamefont
  {Bramante}, \citenamefont {Fukushima}, \citenamefont {Kumar},\ and\
  \citenamefont {Stopnitzky}}]{Bramante:2013nma}%
  \BibitemOpen
  \bibfield  {author} {\bibinfo {author} {\bibfnamefont {J.}~\bibnamefont
  {Bramante}}, \bibinfo {author} {\bibfnamefont {K.}~\bibnamefont {Fukushima}},
  \bibinfo {author} {\bibfnamefont {J.}~\bibnamefont {Kumar}},\ and\ \bibinfo
  {author} {\bibfnamefont {E.}~\bibnamefont {Stopnitzky}},\ }\bibfield  {title}
  {\bibinfo {title} {{Bounds on self-interacting fermion dark matter from
  observations of old neutron stars}},\ }\href
  {https://doi.org/10.1103/PhysRevD.89.015010} {\bibfield  {journal} {\bibinfo
  {journal} {Phys. Rev. D}\ }\textbf {\bibinfo {volume} {89}},\ \bibinfo
  {pages} {015010} (\bibinfo {year} {2014})},\ \Eprint
  {https://arxiv.org/abs/1310.3509} {arXiv:1310.3509 [hep-ph]} \BibitemShut
  {NoStop}%
\bibitem [{\citenamefont {Zhang}\ \emph {et~al.}(2024)\citenamefont {Zhang},
  \citenamefont {Hagimoto},\ and\ \citenamefont {Long}}]{Zhang:2023vva}%
  \BibitemOpen
  \bibfield  {author} {\bibinfo {author} {\bibfnamefont {H.-Y.}\ \bibnamefont
  {Zhang}}, \bibinfo {author} {\bibfnamefont {R.}~\bibnamefont {Hagimoto}},\
  and\ \bibinfo {author} {\bibfnamefont {A.~J.}\ \bibnamefont {Long}},\
  }\bibfield  {title} {\bibinfo {title} {{Neutron star cooling with
  lepton-flavor-violating axions}},\ }\href
  {https://doi.org/10.1103/PhysRevD.109.103005} {\bibfield  {journal} {\bibinfo
   {journal} {Phys. Rev. D}\ }\textbf {\bibinfo {volume} {109}},\ \bibinfo
  {pages} {103005} (\bibinfo {year} {2024})},\ \Eprint
  {https://arxiv.org/abs/2309.03889} {arXiv:2309.03889 [hep-ph]} \BibitemShut
  {NoStop}%
\bibitem [{\citenamefont {Raffelt}(1996)}]{Raffelt:1996wa}%
  \BibitemOpen
  \bibfield  {author} {\bibinfo {author} {\bibfnamefont {G.~G.}\ \bibnamefont
  {Raffelt}},\ }\href@noop {} {\emph {\bibinfo {title} {{Stars as laboratories
  for fundamental physics}: {The astrophysics of neutrinos, axions, and other
  weakly interacting particles}}}}\ (\bibinfo {year} {1996})\BibitemShut
  {NoStop}%
\bibitem [{\citenamefont {Walters}\ \emph {et~al.}(2024)\citenamefont
  {Walters}, \citenamefont {Shroyer}, \citenamefont {Edenton}, \citenamefont
  {Agrawal}, \citenamefont {Johnson}, \citenamefont {Kavanagh}, \citenamefont
  {Marsh},\ and\ \citenamefont {Visinelli}}]{Walters:2024vaw}%
  \BibitemOpen
  \bibfield  {author} {\bibinfo {author} {\bibfnamefont {L.}~\bibnamefont
  {Walters}}, \bibinfo {author} {\bibfnamefont {J.~E.}\ \bibnamefont
  {Shroyer}}, \bibinfo {author} {\bibfnamefont {M.}~\bibnamefont {Edenton}},
  \bibinfo {author} {\bibfnamefont {P.}~\bibnamefont {Agrawal}}, \bibinfo
  {author} {\bibfnamefont {B.}~\bibnamefont {Johnson}}, \bibinfo {author}
  {\bibfnamefont {B.~J.}\ \bibnamefont {Kavanagh}}, \bibinfo {author}
  {\bibfnamefont {D.~J.~E.}\ \bibnamefont {Marsh}},\ and\ \bibinfo {author}
  {\bibfnamefont {L.}~\bibnamefont {Visinelli}},\ }\bibfield  {title} {\bibinfo
  {title} {{Axions in Andromeda: Searching for minicluster-neutron star
  encounters with the Green Bank Telescope}},\ }\href
  {https://doi.org/10.1103/PhysRevD.110.123002} {\bibfield  {journal} {\bibinfo
   {journal} {Phys. Rev. D}\ }\textbf {\bibinfo {volume} {110}},\ \bibinfo
  {pages} {123002} (\bibinfo {year} {2024})},\ \Eprint
  {https://arxiv.org/abs/2407.13060} {arXiv:2407.13060 [astro-ph.CO]}
  \BibitemShut {NoStop}%
\bibitem [{\citenamefont {Edwards}\ \emph {et~al.}(2021)\citenamefont
  {Edwards}, \citenamefont {Kavanagh}, \citenamefont {Visinelli},\ and\
  \citenamefont {Weniger}}]{Edwards:2020afl}%
  \BibitemOpen
  \bibfield  {author} {\bibinfo {author} {\bibfnamefont {T.~D.~P.}\
  \bibnamefont {Edwards}}, \bibinfo {author} {\bibfnamefont {B.~J.}\
  \bibnamefont {Kavanagh}}, \bibinfo {author} {\bibfnamefont {L.}~\bibnamefont
  {Visinelli}},\ and\ \bibinfo {author} {\bibfnamefont {C.}~\bibnamefont
  {Weniger}},\ }\bibfield  {title} {\bibinfo {title} {{Transient Radio
  Signatures from Neutron Star Encounters with QCD Axion Miniclusters}},\
  }\href {https://doi.org/10.1103/PhysRevLett.127.131103} {\bibfield  {journal}
  {\bibinfo  {journal} {Phys. Rev. Lett.}\ }\textbf {\bibinfo {volume} {127}},\
  \bibinfo {pages} {131103} (\bibinfo {year} {2021})},\ \Eprint
  {https://arxiv.org/abs/2011.05378} {arXiv:2011.05378 [hep-ph]} \BibitemShut
  {NoStop}%
\bibitem [{\citenamefont {Zhang}(2023)}]{Zhang:2023ktk}%
  \BibitemOpen
  \bibfield  {author} {\bibinfo {author} {\bibfnamefont {H.-Y.}\ \bibnamefont
  {Zhang}},\ }\emph {\bibinfo {title} {{Probing ultralight dark fields in
  cosmological and astrophysical systems}}},\ \href@noop {} {Ph.D. thesis},\
  \bibinfo  {school} {Rice U.} (\bibinfo {year} {2023}),\ \Eprint
  {https://arxiv.org/abs/2401.00043} {arXiv:2401.00043 [hep-ph]} \BibitemShut
  {NoStop}%
\bibitem [{\citenamefont {Yin}\ \emph {et~al.}(2026)\citenamefont {Yin},
  \citenamefont {Balaji}, \citenamefont {Fairbairn},\ and\ \citenamefont
  {Marsh}}]{Yin:2026rfe}%
  \BibitemOpen
  \bibfield  {author} {\bibinfo {author} {\bibfnamefont {Z.}~\bibnamefont
  {Yin}}, \bibinfo {author} {\bibfnamefont {S.}~\bibnamefont {Balaji}},
  \bibinfo {author} {\bibfnamefont {M.}~\bibnamefont {Fairbairn}},\ and\
  \bibinfo {author} {\bibfnamefont {D.~J.~E.}\ \bibnamefont {Marsh}},\
  }\bibfield  {title} {\bibinfo {title} {{Constraining axion quadratic
  couplings with the Hulse-Taylor binary system}},\ }\href@noop {} {\
  (\bibinfo {year} {2026})},\ \Eprint {https://arxiv.org/abs/2607.25631}
  {arXiv:2607.25631 [hep-ph]} \BibitemShut {NoStop}%
\bibitem [{\citenamefont {Herrero}\ \emph {et~al.}(2019)\citenamefont
  {Herrero}, \citenamefont {P{\'e}rez-Garc{\'\i}a}, \citenamefont {Silk},\ and\
  \citenamefont {Albertus}}]{Herrero:2019esf}%
  \BibitemOpen
  \bibfield  {author} {\bibinfo {author} {\bibfnamefont {A.}~\bibnamefont
  {Herrero}}, \bibinfo {author} {\bibfnamefont {M.~A.}\ \bibnamefont
  {P{\'e}rez-Garc{\'\i}a}}, \bibinfo {author} {\bibfnamefont {J.}~\bibnamefont
  {Silk}},\ and\ \bibinfo {author} {\bibfnamefont {C.}~\bibnamefont
  {Albertus}},\ }\bibfield  {title} {\bibinfo {title} {{Dark matter and bubble
  nucleation in old neutron stars}},\ }\href
  {https://doi.org/10.1103/PhysRevD.100.103019} {\bibfield  {journal} {\bibinfo
   {journal} {Phys. Rev. D}\ }\textbf {\bibinfo {volume} {100}},\ \bibinfo
  {pages} {103019} (\bibinfo {year} {2019})},\ \Eprint
  {https://arxiv.org/abs/1905.00893} {arXiv:1905.00893 [hep-ph]} \BibitemShut
  {NoStop}%
\bibitem [{\citenamefont {{\'A}ngeles P{\'e}rez-Garc{\'\i}a}\ and\
  \citenamefont {Silk}(2015)}]{AngelesPerez-Garcia:2014cho}%
  \BibitemOpen
  \bibfield  {author} {\bibinfo {author} {\bibfnamefont {M.}~\bibnamefont
  {{\'A}ngeles P{\'e}rez-Garc{\'\i}a}}\ and\ \bibinfo {author} {\bibfnamefont
  {J.}~\bibnamefont {Silk}},\ }\bibfield  {title} {\bibinfo {title}
  {{Constraining decaying dark matter with neutron stars}},\ }\href
  {https://doi.org/10.1016/j.physletb.2015.03.026} {\bibfield  {journal}
  {\bibinfo  {journal} {Phys. Lett. B}\ }\textbf {\bibinfo {volume} {744}},\
  \bibinfo {pages} {13} (\bibinfo {year} {2015})},\ \Eprint
  {https://arxiv.org/abs/1403.6111} {arXiv:1403.6111 [astro-ph.SR]}
  \BibitemShut {NoStop}%
\bibitem [{\citenamefont {Bhutani}\ \emph {et~al.}(2026)\citenamefont
  {Bhutani}, \citenamefont {Raj},\ and\ \citenamefont
  {Zuraiq}}]{Bhutani:2025jfo}%
  \BibitemOpen
  \bibfield  {author} {\bibinfo {author} {\bibfnamefont {A.}~\bibnamefont
  {Bhutani}}, \bibinfo {author} {\bibfnamefont {N.}~\bibnamefont {Raj}},\ and\
  \bibinfo {author} {\bibfnamefont {Z.}~\bibnamefont {Zuraiq}},\ }\bibfield
  {title} {\bibinfo {title} {{Dark, deep, deconfining: Phase transitions in
  neutron stars as powerful probes of hidden sectors}},\ }\href
  {https://doi.org/10.1103/yblk-411t} {\bibfield  {journal} {\bibinfo
  {journal} {Phys. Rev. D}\ }\textbf {\bibinfo {volume} {113}},\ \bibinfo
  {pages} {103046} (\bibinfo {year} {2026})},\ \Eprint
  {https://arxiv.org/abs/2507.08076} {arXiv:2507.08076 [hep-ph]} \BibitemShut
  {NoStop}%
\bibitem [{\citenamefont {Mandhai}\ \emph {et~al.}(2018)\citenamefont
  {Mandhai}, \citenamefont {Tanvir}, \citenamefont {Lamb}, \citenamefont
  {Levan},\ and\ \citenamefont {Tsang}}]{Mandhai:2018cdl}%
  \BibitemOpen
  \bibfield  {author} {\bibinfo {author} {\bibfnamefont {S.}~\bibnamefont
  {Mandhai}}, \bibinfo {author} {\bibfnamefont {N.}~\bibnamefont {Tanvir}},
  \bibinfo {author} {\bibfnamefont {G.}~\bibnamefont {Lamb}}, \bibinfo {author}
  {\bibfnamefont {A.}~\bibnamefont {Levan}},\ and\ \bibinfo {author}
  {\bibfnamefont {D.}~\bibnamefont {Tsang}},\ }\bibfield  {title} {\bibinfo
  {title} {{The Rate of Short-Duration Gamma-Ray Bursts in the Local
  Universe}},\ }\href {https://doi.org/10.3390/galaxies6040130} {\bibfield
  {journal} {\bibinfo  {journal} {Galaxies}\ }\textbf {\bibinfo {volume} {6}},\
  \bibinfo {pages} {130} (\bibinfo {year} {2018})},\ \Eprint
  {https://arxiv.org/abs/1812.00507} {arXiv:1812.00507 [astro-ph.HE]}
  \BibitemShut {NoStop}%
\bibitem [{\citenamefont {Dichiara}\ \emph {et~al.}(2020)\citenamefont
  {Dichiara}, \citenamefont {Troja}, \citenamefont {O'Connor}, \citenamefont
  {Marshall}, \citenamefont {Beniamini}, \citenamefont {Cannizzo},
  \citenamefont {Lien},\ and\ \citenamefont {Sakamoto}}]{Dichiara:2019kuw}%
  \BibitemOpen
  \bibfield  {author} {\bibinfo {author} {\bibfnamefont {S.}~\bibnamefont
  {Dichiara}}, \bibinfo {author} {\bibfnamefont {E.}~\bibnamefont {Troja}},
  \bibinfo {author} {\bibfnamefont {B.}~\bibnamefont {O'Connor}}, \bibinfo
  {author} {\bibfnamefont {F.~E.}\ \bibnamefont {Marshall}}, \bibinfo {author}
  {\bibfnamefont {P.}~\bibnamefont {Beniamini}}, \bibinfo {author}
  {\bibfnamefont {J.~K.}\ \bibnamefont {Cannizzo}}, \bibinfo {author}
  {\bibfnamefont {A.~Y.}\ \bibnamefont {Lien}},\ and\ \bibinfo {author}
  {\bibfnamefont {T.}~\bibnamefont {Sakamoto}},\ }\bibfield  {title} {\bibinfo
  {title} {{Short gamma-ray bursts within 200 Mpc}},\ }\href
  {https://doi.org/10.1093/mnras/staa124} {\bibfield  {journal} {\bibinfo
  {journal} {Mon. Not. Roy. Astron. Soc.}\ }\textbf {\bibinfo {volume} {492}},\
  \bibinfo {pages} {5011} (\bibinfo {year} {2020})},\ \Eprint
  {https://arxiv.org/abs/1912.08698} {arXiv:1912.08698 [astro-ph.HE]}
  \BibitemShut {NoStop}%
\bibitem [{\citenamefont {Escorial}\ \emph {et~al.}(2023)\citenamefont
  {Escorial} \emph {et~al.}}]{Escorial:2022nvp}%
  \BibitemOpen
  \bibfield  {author} {\bibinfo {author} {\bibfnamefont {A.~R.}\ \bibnamefont
  {Escorial}} \emph {et~al.},\ }\bibfield  {title} {\bibinfo {title} {{The Jet
  Opening Angle and Event Rate Distributions of Short Gamma-Ray Bursts from
  Late-time X-Ray Afterglows}},\ }\href
  {https://doi.org/10.3847/1538-4357/acf830} {\bibfield  {journal} {\bibinfo
  {journal} {Astrophys. J.}\ }\textbf {\bibinfo {volume} {959}},\ \bibinfo
  {pages} {13} (\bibinfo {year} {2023})},\ \Eprint
  {https://arxiv.org/abs/2210.05695} {arXiv:2210.05695 [astro-ph.HE]}
  \BibitemShut {NoStop}%
\bibitem [{\citenamefont {Fong}\ \emph {et~al.}(2015)\citenamefont {Fong},
  \citenamefont {Berger}, \citenamefont {Margutti},\ and\ \citenamefont
  {Zauderer}}]{Fong:2015oha}%
  \BibitemOpen
  \bibfield  {author} {\bibinfo {author} {\bibfnamefont {W.-f.}\ \bibnamefont
  {Fong}}, \bibinfo {author} {\bibfnamefont {E.}~\bibnamefont {Berger}},
  \bibinfo {author} {\bibfnamefont {R.}~\bibnamefont {Margutti}},\ and\
  \bibinfo {author} {\bibfnamefont {B.~A.}\ \bibnamefont {Zauderer}},\
  }\bibfield  {title} {\bibinfo {title} {{A Decade of Short-duration Gamma-ray
  Burst Broadband Afterglows: Energetics, Circumburst Densities, and jet
  Opening Angles}},\ }\href {https://doi.org/10.1088/0004-637X/815/2/102}
  {\bibfield  {journal} {\bibinfo  {journal} {Astrophys. J.}\ }\textbf
  {\bibinfo {volume} {815}},\ \bibinfo {pages} {102} (\bibinfo {year}
  {2015})},\ \Eprint {https://arxiv.org/abs/1509.02922} {arXiv:1509.02922
  [astro-ph.HE]} \BibitemShut {NoStop}%
\bibitem [{\citenamefont {Fukugita}\ and\ \citenamefont
  {Peebles}(2004)}]{Fukugita:2004ee}%
  \BibitemOpen
  \bibfield  {author} {\bibinfo {author} {\bibfnamefont {M.}~\bibnamefont
  {Fukugita}}\ and\ \bibinfo {author} {\bibfnamefont {P.~J.~E.}\ \bibnamefont
  {Peebles}},\ }\bibfield  {title} {\bibinfo {title} {{The Cosmic energy
  inventory}},\ }\href {https://doi.org/10.1086/425155} {\bibfield  {journal}
  {\bibinfo  {journal} {Astrophys. J.}\ }\textbf {\bibinfo {volume} {616}},\
  \bibinfo {pages} {643} (\bibinfo {year} {2004})},\ \Eprint
  {https://arxiv.org/abs/astro-ph/0406095} {arXiv:astro-ph/0406095}
  \BibitemShut {NoStop}%
\bibitem [{\citenamefont {Elbert}\ \emph {et~al.}(2018)\citenamefont {Elbert},
  \citenamefont {Bullock},\ and\ \citenamefont {Kaplinghat}}]{Elbert:2017sbr}%
  \BibitemOpen
  \bibfield  {author} {\bibinfo {author} {\bibfnamefont {O.~D.}\ \bibnamefont
  {Elbert}}, \bibinfo {author} {\bibfnamefont {J.~S.}\ \bibnamefont
  {Bullock}},\ and\ \bibinfo {author} {\bibfnamefont {M.}~\bibnamefont
  {Kaplinghat}},\ }\bibfield  {title} {\bibinfo {title} {{Counting Black Holes:
  The Cosmic Stellar Remnant Population and Implications for LIGO}},\ }\href
  {https://doi.org/10.1093/mnras/stx1959} {\bibfield  {journal} {\bibinfo
  {journal} {Mon. Not. Roy. Astron. Soc.}\ }\textbf {\bibinfo {volume} {473}},\
  \bibinfo {pages} {1186} (\bibinfo {year} {2018})},\ \Eprint
  {https://arxiv.org/abs/1703.02551} {arXiv:1703.02551 [astro-ph.GA]}
  \BibitemShut {NoStop}%
\bibitem [{\citenamefont {Maraston}\ \emph {et~al.}(2025)\citenamefont
  {Maraston}, \citenamefont {Limongi}, \citenamefont {Neumann}, \citenamefont
  {Roberti}, \citenamefont {Chieffi}, \citenamefont {Thomas},\ and\
  \citenamefont {Lian}}]{2025MNRAS.540.2359M}%
  \BibitemOpen
  \bibfield  {author} {\bibinfo {author} {\bibfnamefont {C.}~\bibnamefont
  {Maraston}}, \bibinfo {author} {\bibfnamefont {M.}~\bibnamefont {Limongi}},
  \bibinfo {author} {\bibfnamefont {J.}~\bibnamefont {Neumann}}, \bibinfo
  {author} {\bibfnamefont {L.}~\bibnamefont {Roberti}}, \bibinfo {author}
  {\bibfnamefont {A.}~\bibnamefont {Chieffi}}, \bibinfo {author} {\bibfnamefont
  {D.}~\bibnamefont {Thomas}},\ and\ \bibinfo {author} {\bibfnamefont
  {J.}~\bibnamefont {Lian}},\ }\bibfield  {title} {\bibinfo {title} {Stellar
  population modelling of neutron stars and black holes: spatially resolved
  graveyards in {MaNGA}/{SDSS-IV} galaxies},\ }\href
  {https://doi.org/10.1093/mnras/staf801} {\bibfield  {journal} {\bibinfo
  {journal} {Monthly Notices of the Royal Astronomical Society}\ }\textbf
  {\bibinfo {volume} {540}},\ \bibinfo {pages} {2359} (\bibinfo {year}
  {2025})},\ \Eprint {https://arxiv.org/abs/2505.15691} {arXiv:2505.15691
  [astro-ph.GA]} \BibitemShut {NoStop}%
\bibitem [{\citenamefont {Abadie}\ \emph {et~al.}(2010)\citenamefont {Abadie}
  \emph {et~al.}}]{LIGOScientific:2010nhs}%
  \BibitemOpen
  \bibfield  {author} {\bibinfo {author} {\bibfnamefont {J.}~\bibnamefont
  {Abadie}} \emph {et~al.} (\bibinfo {collaboration} {LIGO Scientific,
  VIRGO}),\ }\bibfield  {title} {\bibinfo {title} {{Predictions for the Rates
  of Compact Binary Coalescences Observable by Ground-based Gravitational-wave
  Detectors}},\ }\href {https://doi.org/10.1088/0264-9381/27/17/173001}
  {\bibfield  {journal} {\bibinfo  {journal} {Class. Quant. Grav.}\ }\textbf
  {\bibinfo {volume} {27}},\ \bibinfo {pages} {173001} (\bibinfo {year}
  {2010})},\ \Eprint {https://arxiv.org/abs/1003.2480} {arXiv:1003.2480
  [astro-ph.HE]} \BibitemShut {NoStop}%
\bibitem [{\citenamefont {Ofek}(2009)}]{2009PASP..121..814O}%
  \BibitemOpen
  \bibfield  {author} {\bibinfo {author} {\bibfnamefont {E.~O.}\ \bibnamefont
  {Ofek}},\ }\bibfield  {title} {\bibinfo {title} {{Space and velocity
  distributions of Galactic isolated old Neutron stars}},\ }\href
  {https://doi.org/10.1086/605389} {\bibfield  {journal} {\bibinfo  {journal}
  {Publ. Astron. Soc. Pac.}\ }\textbf {\bibinfo {volume} {121}},\ \bibinfo
  {pages} {814} (\bibinfo {year} {2009})},\ \Eprint
  {https://arxiv.org/abs/0910.3684} {arXiv:0910.3684 [astro-ph.GA]}
  \BibitemShut {NoStop}%
\bibitem [{\citenamefont {Sartore}\ \emph {et~al.}(2010)\citenamefont
  {Sartore}, \citenamefont {Ripamonti}, \citenamefont {Treves},\ and\
  \citenamefont {Turolla}}]{2010A&A...510A..23S}%
  \BibitemOpen
  \bibfield  {author} {\bibinfo {author} {\bibfnamefont {N.}~\bibnamefont
  {Sartore}}, \bibinfo {author} {\bibfnamefont {E.}~\bibnamefont {Ripamonti}},
  \bibinfo {author} {\bibfnamefont {A.}~\bibnamefont {Treves}},\ and\ \bibinfo
  {author} {\bibfnamefont {R.}~\bibnamefont {Turolla}},\ }\bibfield  {title}
  {\bibinfo {title} {{Galactic neutron stars I. Space and velocity
  distributions in the disk and in the halo}},\ }\href
  {https://doi.org/10.1051/0004-6361/200912222} {\bibfield  {journal} {\bibinfo
   {journal} {Astron. Astrophys.}\ }\textbf {\bibinfo {volume} {510}},\
  \bibinfo {pages} {A23} (\bibinfo {year} {2010})},\ \Eprint
  {https://arxiv.org/abs/0908.3182} {arXiv:0908.3182 [astro-ph.GA]}
  \BibitemShut {NoStop}%
\bibitem [{\citenamefont {You}\ \emph {et~al.}(2025)\citenamefont {You} \emph
  {et~al.}}]{You:2024bmk}%
  \BibitemOpen
  \bibfield  {author} {\bibinfo {author} {\bibfnamefont {Z.-Q.}\ \bibnamefont
  {You}} \emph {et~al.},\ }\bibfield  {title} {\bibinfo {title} {{Determination
  of the birth-mass function of neutron stars from observations}},\ }\href
  {https://doi.org/10.1038/s41550-025-02487-w} {\bibfield  {journal} {\bibinfo
  {journal} {Nature Astron.}\ }\textbf {\bibinfo {volume} {9}},\ \bibinfo
  {pages} {552} (\bibinfo {year} {2025})},\ \Eprint
  {https://arxiv.org/abs/2412.05524} {arXiv:2412.05524 [astro-ph.HE]}
  \BibitemShut {NoStop}%
\bibitem [{\citenamefont {Bromley}(2011)}]{Bromley:2011aa}%
  \BibitemOpen
  \bibfield  {author} {\bibinfo {author} {\bibfnamefont {B.~C.}\ \bibnamefont
  {Bromley}},\ }\bibfield  {title} {\bibinfo {title} {{Gravitationally Focused
  Dark Matter Around Compact Stars}},\ }\href
  {https://doi.org/10.1088/0067-0049/197/2/37} {\bibfield  {journal} {\bibinfo
  {journal} {Astrophys. J. Suppl.}\ }\textbf {\bibinfo {volume} {197}},\
  \bibinfo {pages} {37} (\bibinfo {year} {2011})},\ \Eprint
  {https://arxiv.org/abs/1112.2355} {arXiv:1112.2355 [astro-ph.HE]}
  \BibitemShut {NoStop}%
\bibitem [{\citenamefont {McDaniel}\ \emph {et~al.}(2024)\citenamefont
  {McDaniel}, \citenamefont {Ajello}, \citenamefont {Karwin}, \citenamefont
  {Di~Mauro}, \citenamefont {Drlica-Wagner},\ and\ \citenamefont
  {S{\'a}nchez-Conde}}]{McDaniel:2023bju}%
  \BibitemOpen
  \bibfield  {author} {\bibinfo {author} {\bibfnamefont {A.}~\bibnamefont
  {McDaniel}}, \bibinfo {author} {\bibfnamefont {M.}~\bibnamefont {Ajello}},
  \bibinfo {author} {\bibfnamefont {C.~M.}\ \bibnamefont {Karwin}}, \bibinfo
  {author} {\bibfnamefont {M.}~\bibnamefont {Di~Mauro}}, \bibinfo {author}
  {\bibfnamefont {A.}~\bibnamefont {Drlica-Wagner}},\ and\ \bibinfo {author}
  {\bibfnamefont {M.~A.}\ \bibnamefont {S{\'a}nchez-Conde}},\ }\bibfield
  {title} {\bibinfo {title} {{Legacy analysis of dark matter annihilation from
  the Milky~Way dwarf spheroidal galaxies with 14~years of Fermi-LAT data}},\
  }\href {https://doi.org/10.1103/PhysRevD.109.063024} {\bibfield  {journal}
  {\bibinfo  {journal} {Phys. Rev. D}\ }\textbf {\bibinfo {volume} {109}},\
  \bibinfo {pages} {063024} (\bibinfo {year} {2024})},\ \Eprint
  {https://arxiv.org/abs/2311.04982} {arXiv:2311.04982 [astro-ph.HE]}
  \BibitemShut {NoStop}%
\bibitem [{\citenamefont {Abdalla}\ \emph {et~al.}(2022)\citenamefont {Abdalla}
  \emph {et~al.}}]{HESS:2022ygk}%
  \BibitemOpen
  \bibfield  {author} {\bibinfo {author} {\bibfnamefont {H.}~\bibnamefont
  {Abdalla}} \emph {et~al.} (\bibinfo {collaboration} {H.E.S.S.}),\ }\bibfield
  {title} {\bibinfo {title} {{Search for Dark Matter Annihilation Signals in
  the H.E.S.S. Inner Galaxy Survey}},\ }\href
  {https://doi.org/10.1103/PhysRevLett.129.111101} {\bibfield  {journal}
  {\bibinfo  {journal} {Phys. Rev. Lett.}\ }\textbf {\bibinfo {volume} {129}},\
  \bibinfo {pages} {111101} (\bibinfo {year} {2022})},\ \Eprint
  {https://arxiv.org/abs/2207.10471} {arXiv:2207.10471 [astro-ph.HE]}
  \BibitemShut {NoStop}%
\bibitem [{\citenamefont {Aghanim}\ \emph {et~al.}(2020)\citenamefont {Aghanim}
  \emph {et~al.}}]{Planck:2018vyg}%
  \BibitemOpen
  \bibfield  {author} {\bibinfo {author} {\bibfnamefont {N.}~\bibnamefont
  {Aghanim}} \emph {et~al.} (\bibinfo {collaboration} {Planck}),\ }\bibfield
  {title} {\bibinfo {title} {{Planck 2018 results. VI. Cosmological
  parameters}},\ }\href {https://doi.org/10.1051/0004-6361/201833910}
  {\bibfield  {journal} {\bibinfo  {journal} {Astron. Astrophys.}\ }\textbf
  {\bibinfo {volume} {641}},\ \bibinfo {pages} {A6} (\bibinfo {year} {2020})},\
  \bibinfo {note} {[Erratum: Astron.Astrophys. 652, C4 (2021)]},\ \Eprint
  {https://arxiv.org/abs/1807.06209} {arXiv:1807.06209 [astro-ph.CO]}
  \BibitemShut {NoStop}%
\bibitem [{\citenamefont {Steigman}\ \emph {et~al.}(2012)\citenamefont
  {Steigman}, \citenamefont {Dasgupta},\ and\ \citenamefont
  {Beacom}}]{Steigman:2012nb}%
  \BibitemOpen
  \bibfield  {author} {\bibinfo {author} {\bibfnamefont {G.}~\bibnamefont
  {Steigman}}, \bibinfo {author} {\bibfnamefont {B.}~\bibnamefont {Dasgupta}},\
  and\ \bibinfo {author} {\bibfnamefont {J.~F.}\ \bibnamefont {Beacom}},\
  }\bibfield  {title} {\bibinfo {title} {{Precise Relic WIMP Abundance and its
  Impact on Searches for Dark Matter Annihilation}},\ }\href
  {https://doi.org/10.1103/PhysRevD.86.023506} {\bibfield  {journal} {\bibinfo
  {journal} {Phys. Rev. D}\ }\textbf {\bibinfo {volume} {86}},\ \bibinfo
  {pages} {023506} (\bibinfo {year} {2012})},\ \Eprint
  {https://arxiv.org/abs/1204.3622} {arXiv:1204.3622 [hep-ph]} \BibitemShut
  {NoStop}%
\bibitem [{\citenamefont {Baum}\ \emph {et~al.}(2017)\citenamefont {Baum},
  \citenamefont {Visinelli}, \citenamefont {Freese},\ and\ \citenamefont
  {Stengel}}]{Baum:2016oow}%
  \BibitemOpen
  \bibfield  {author} {\bibinfo {author} {\bibfnamefont {S.}~\bibnamefont
  {Baum}}, \bibinfo {author} {\bibfnamefont {L.}~\bibnamefont {Visinelli}},
  \bibinfo {author} {\bibfnamefont {K.}~\bibnamefont {Freese}},\ and\ \bibinfo
  {author} {\bibfnamefont {P.}~\bibnamefont {Stengel}},\ }\bibfield  {title}
  {\bibinfo {title} {{Dark matter capture, subdominant WIMPs, and neutrino
  observatories}},\ }\href {https://doi.org/10.1103/PhysRevD.95.043007}
  {\bibfield  {journal} {\bibinfo  {journal} {Phys. Rev. D}\ }\textbf {\bibinfo
  {volume} {95}},\ \bibinfo {pages} {043007} (\bibinfo {year} {2017})},\
  \Eprint {https://arxiv.org/abs/1611.09665} {arXiv:1611.09665 [astro-ph.CO]}
  \BibitemShut {NoStop}%
\bibitem [{\citenamefont {Ando}\ and\ \citenamefont
  {Ishiwata}(2015)}]{Ando:2015qda}%
  \BibitemOpen
  \bibfield  {author} {\bibinfo {author} {\bibfnamefont {S.}~\bibnamefont
  {Ando}}\ and\ \bibinfo {author} {\bibfnamefont {K.}~\bibnamefont
  {Ishiwata}},\ }\bibfield  {title} {\bibinfo {title} {{Constraints on decaying
  dark matter from the extragalactic gamma-ray background}},\ }\href
  {https://doi.org/10.1088/1475-7516/2015/05/024} {\bibfield  {journal}
  {\bibinfo  {journal} {JCAP}\ }\textbf {\bibinfo {volume} {05}},\ \bibinfo
  {pages} {024}},\ \Eprint {https://arxiv.org/abs/1502.02007} {arXiv:1502.02007
  [astro-ph.CO]} \BibitemShut {NoStop}%
\bibitem [{\citenamefont {Baring}\ \emph {et~al.}(2016)\citenamefont {Baring},
  \citenamefont {Ghosh}, \citenamefont {Queiroz},\ and\ \citenamefont
  {Sinha}}]{Baring:2015sza}%
  \BibitemOpen
  \bibfield  {author} {\bibinfo {author} {\bibfnamefont {M.~G.}\ \bibnamefont
  {Baring}}, \bibinfo {author} {\bibfnamefont {T.}~\bibnamefont {Ghosh}},
  \bibinfo {author} {\bibfnamefont {F.~S.}\ \bibnamefont {Queiroz}},\ and\
  \bibinfo {author} {\bibfnamefont {K.}~\bibnamefont {Sinha}},\ }\bibfield
  {title} {\bibinfo {title} {{New Limits on the Dark Matter Lifetime from Dwarf
  Spheroidal Galaxies using Fermi-LAT}},\ }\href
  {https://doi.org/10.1103/PhysRevD.93.103009} {\bibfield  {journal} {\bibinfo
  {journal} {Phys. Rev. D}\ }\textbf {\bibinfo {volume} {93}},\ \bibinfo
  {pages} {103009} (\bibinfo {year} {2016})},\ \Eprint
  {https://arxiv.org/abs/1510.00389} {arXiv:1510.00389 [hep-ph]} \BibitemShut
  {NoStop}%
\bibitem [{\citenamefont {Slatyer}\ and\ \citenamefont
  {Wu}(2017)}]{Slatyer:2016qyl}%
  \BibitemOpen
  \bibfield  {author} {\bibinfo {author} {\bibfnamefont {T.~R.}\ \bibnamefont
  {Slatyer}}\ and\ \bibinfo {author} {\bibfnamefont {C.-L.}\ \bibnamefont
  {Wu}},\ }\bibfield  {title} {\bibinfo {title} {{General Constraints on Dark
  Matter Decay from the Cosmic Microwave Background}},\ }\href
  {https://doi.org/10.1103/PhysRevD.95.023010} {\bibfield  {journal} {\bibinfo
  {journal} {Phys. Rev. D}\ }\textbf {\bibinfo {volume} {95}},\ \bibinfo
  {pages} {023010} (\bibinfo {year} {2017})},\ \Eprint
  {https://arxiv.org/abs/1610.06933} {arXiv:1610.06933 [astro-ph.CO]}
  \BibitemShut {NoStop}%
\bibitem [{\citenamefont {Jacobs}\ \emph {et~al.}(2015)\citenamefont {Jacobs},
  \citenamefont {Starkman},\ and\ \citenamefont {Lynn}}]{Jacobs:2014yca}%
  \BibitemOpen
  \bibfield  {author} {\bibinfo {author} {\bibfnamefont {D.~M.}\ \bibnamefont
  {Jacobs}}, \bibinfo {author} {\bibfnamefont {G.~D.}\ \bibnamefont
  {Starkman}},\ and\ \bibinfo {author} {\bibfnamefont {B.~W.}\ \bibnamefont
  {Lynn}},\ }\bibfield  {title} {\bibinfo {title} {{Macro Dark Matter}},\
  }\href {https://doi.org/10.1093/mnras/stv774} {\bibfield  {journal} {\bibinfo
   {journal} {Mon. Not. Roy. Astron. Soc.}\ }\textbf {\bibinfo {volume}
  {450}},\ \bibinfo {pages} {3418} (\bibinfo {year} {2015})},\ \Eprint
  {https://arxiv.org/abs/1410.2236} {arXiv:1410.2236 [astro-ph.CO]}
  \BibitemShut {NoStop}%
\bibitem [{\citenamefont {Singh~Sidhu}\ and\ \citenamefont
  {Starkman}(2020)}]{SinghSidhu:2019tbr}%
  \BibitemOpen
  \bibfield  {author} {\bibinfo {author} {\bibfnamefont {J.}~\bibnamefont
  {Singh~Sidhu}}\ and\ \bibinfo {author} {\bibfnamefont {G.~D.}\ \bibnamefont
  {Starkman}},\ }\bibfield  {title} {\bibinfo {title} {{Reconsidering
  astrophysical constraints on macroscopic dark matter}},\ }\href
  {https://doi.org/10.1103/PhysRevD.101.083503} {\bibfield  {journal} {\bibinfo
   {journal} {Phys. Rev. D}\ }\textbf {\bibinfo {volume} {101}},\ \bibinfo
  {pages} {083503} (\bibinfo {year} {2020})},\ \Eprint
  {https://arxiv.org/abs/1912.04053} {arXiv:1912.04053 [astro-ph.CO]}
  \BibitemShut {NoStop}%
\bibitem [{\citenamefont {Bramante}(2026)}]{Bramante:2026wzh}%
  \BibitemOpen
  \bibfield  {author} {\bibinfo {author} {\bibfnamefont {J.}~\bibnamefont
  {Bramante}},\ }\bibfield  {title} {\bibinfo {title} {{Very Heavy and
  Composite Dark Matter: Theory and Experimental Searches}},\ }\href@noop {} {\
   (\bibinfo {year} {2026})},\ \Eprint {https://arxiv.org/abs/2602.23708}
  {arXiv:2602.23708 [hep-ph]} \BibitemShut {NoStop}%
\bibitem [{\citenamefont {Graham}\ \emph {et~al.}(2018)\citenamefont {Graham},
  \citenamefont {Janish}, \citenamefont {Narayan}, \citenamefont {Rajendran},\
  and\ \citenamefont {Riggins}}]{Graham:2018efk}%
  \BibitemOpen
  \bibfield  {author} {\bibinfo {author} {\bibfnamefont {P.~W.}\ \bibnamefont
  {Graham}}, \bibinfo {author} {\bibfnamefont {R.}~\bibnamefont {Janish}},
  \bibinfo {author} {\bibfnamefont {V.}~\bibnamefont {Narayan}}, \bibinfo
  {author} {\bibfnamefont {S.}~\bibnamefont {Rajendran}},\ and\ \bibinfo
  {author} {\bibfnamefont {P.}~\bibnamefont {Riggins}},\ }\bibfield  {title}
  {\bibinfo {title} {{White Dwarfs as Dark Matter Detectors}},\ }\href
  {https://doi.org/10.1103/PhysRevD.98.115027} {\bibfield  {journal} {\bibinfo
  {journal} {Phys. Rev. D}\ }\textbf {\bibinfo {volume} {98}},\ \bibinfo
  {pages} {115027} (\bibinfo {year} {2018})},\ \Eprint
  {https://arxiv.org/abs/1805.07381} {arXiv:1805.07381 [hep-ph]} \BibitemShut
  {NoStop}%
\bibitem [{\citenamefont {Zhang}\ \emph {et~al.}(2020)\citenamefont {Zhang},
  \citenamefont {Amin}, \citenamefont {Copeland}, \citenamefont {Saffin},\ and\
  \citenamefont {Lozanov}}]{Zhang:2020bec}%
  \BibitemOpen
  \bibfield  {author} {\bibinfo {author} {\bibfnamefont {H.-Y.}\ \bibnamefont
  {Zhang}}, \bibinfo {author} {\bibfnamefont {M.~A.}\ \bibnamefont {Amin}},
  \bibinfo {author} {\bibfnamefont {E.~J.}\ \bibnamefont {Copeland}}, \bibinfo
  {author} {\bibfnamefont {P.~M.}\ \bibnamefont {Saffin}},\ and\ \bibinfo
  {author} {\bibfnamefont {K.~D.}\ \bibnamefont {Lozanov}},\ }\bibfield
  {title} {\bibinfo {title} {{Classical Decay Rates of Oscillons}},\ }\href
  {https://doi.org/10.1088/1475-7516/2020/07/055} {\bibfield  {journal}
  {\bibinfo  {journal} {JCAP}\ }\textbf {\bibinfo {volume} {07}},\ \bibinfo
  {pages} {055}},\ \Eprint {https://arxiv.org/abs/2004.01202} {arXiv:2004.01202
  [hep-th]} \BibitemShut {NoStop}%
\bibitem [{\citenamefont {Zhang}\ \emph {et~al.}(2022)\citenamefont {Zhang},
  \citenamefont {Jain},\ and\ \citenamefont {Amin}}]{Zhang:2021xxa}%
  \BibitemOpen
  \bibfield  {author} {\bibinfo {author} {\bibfnamefont {H.-Y.}\ \bibnamefont
  {Zhang}}, \bibinfo {author} {\bibfnamefont {M.}~\bibnamefont {Jain}},\ and\
  \bibinfo {author} {\bibfnamefont {M.~A.}\ \bibnamefont {Amin}},\ }\bibfield
  {title} {\bibinfo {title} {{Polarized vector oscillons}},\ }\href
  {https://doi.org/10.1103/PhysRevD.105.096037} {\bibfield  {journal} {\bibinfo
   {journal} {Phys. Rev. D}\ }\textbf {\bibinfo {volume} {105}},\ \bibinfo
  {pages} {096037} (\bibinfo {year} {2022})},\ \Eprint
  {https://arxiv.org/abs/2111.08700} {arXiv:2111.08700 [astro-ph.CO]}
  \BibitemShut {NoStop}%
\bibitem [{\citenamefont {Visinelli}(2021)}]{Visinelli:2021uve}%
  \BibitemOpen
  \bibfield  {author} {\bibinfo {author} {\bibfnamefont {L.}~\bibnamefont
  {Visinelli}},\ }\bibfield  {title} {\bibinfo {title} {{Boson stars and
  oscillatons: A review}},\ }\href {https://doi.org/10.1142/S0218271821300068}
  {\bibfield  {journal} {\bibinfo  {journal} {Int. J. Mod. Phys. D}\ }\textbf
  {\bibinfo {volume} {30}},\ \bibinfo {pages} {2130006} (\bibinfo {year}
  {2021})},\ \Eprint {https://arxiv.org/abs/2109.05481} {arXiv:2109.05481
  [gr-qc]} \BibitemShut {NoStop}%
\bibitem [{\citenamefont {Zhang}(2021)}]{Zhang:2020ntm}%
  \BibitemOpen
  \bibfield  {author} {\bibinfo {author} {\bibfnamefont {H.-Y.}\ \bibnamefont
  {Zhang}},\ }\bibfield  {title} {\bibinfo {title} {{Gravitational effects on
  oscillon lifetimes}},\ }\href {https://doi.org/10.1088/1475-7516/2021/03/102}
  {\bibfield  {journal} {\bibinfo  {journal} {JCAP}\ }\textbf {\bibinfo
  {volume} {03}},\ \bibinfo {pages} {102}},\ \Eprint
  {https://arxiv.org/abs/2011.11720} {arXiv:2011.11720 [hep-th]} \BibitemShut
  {NoStop}%
\bibitem [{\citenamefont {Yin}\ and\ \citenamefont
  {Visinelli}(2024)}]{Yin:2024xov}%
  \BibitemOpen
  \bibfield  {author} {\bibinfo {author} {\bibfnamefont {Z.}~\bibnamefont
  {Yin}}\ and\ \bibinfo {author} {\bibfnamefont {L.}~\bibnamefont
  {Visinelli}},\ }\bibfield  {title} {\bibinfo {title} {{Axion star
  condensation around primordial black holes and microlensing limits}},\ }\href
  {https://doi.org/10.1088/1475-7516/2024/10/013} {\bibfield  {journal}
  {\bibinfo  {journal} {JCAP}\ }\textbf {\bibinfo {volume} {10}},\ \bibinfo
  {pages} {013}},\ \Eprint {https://arxiv.org/abs/2404.10340} {arXiv:2404.10340
  [hep-ph]} \BibitemShut {NoStop}%
\bibitem [{\citenamefont {Zhang}(2025)}]{Zhang:2024bjo}%
  \BibitemOpen
  \bibfield  {author} {\bibinfo {author} {\bibfnamefont {H.-Y.}\ \bibnamefont
  {Zhang}},\ }\bibfield  {title} {\bibinfo {title} {{Unified view of scalar and
  vector dark matter solitons}},\ }\href
  {https://doi.org/10.1007/JHEP04(2025)174} {\bibfield  {journal} {\bibinfo
  {journal} {JHEP}\ }\textbf {\bibinfo {volume} {04}},\ \bibinfo {pages}
  {174}},\ \Eprint {https://arxiv.org/abs/2406.05031} {arXiv:2406.05031
  [hep-ph]} \BibitemShut {NoStop}%
\end{thebibliography}%

\clearpage
\setcounter{section}{0}
\setcounter{equation}{0}
\renewcommand{\theequation}{S\arabic{equation}}

\onecolumngrid
\begin{center}
\textbf{\large {\it Supplemental material on the article:}\\
Metastable Neutron Stars as Transient-Rate Detectors of Heavy Dark Matter}
\end{center}
\twocolumngrid

\section{Population inputs and rate boundaries}
\label{sec:rate_boundary}

The main text gives the present transition rate, Eq.~\eqref{Rtr_benchmark}, and its benchmark exclusion, Eq.~\eqref{Gamma_chi_constraint}. Here we summarize the inputs that are not directly measured and give the exact dependence of the two rate boundaries.

Cosmic remnant inventories imply a present-day NS number density $n_\mrm{NS}\simeq(2.5$--$5)\times10^6\,\mrm{Mpc}^{-3}$, with an overall systematic uncertainty of order a factor of two from stellar population modeling \cite{Fukugita:2004ee,Elbert:2017sbr,2025MNRAS.540.2359M}. Thus we use $n_\mrm{NS}=3\times10^{15}\,\mrm{Gpc}^{-3}$. The metastable fraction is more model dependent. Nucleation calculations typically find a mass window $\Delta M_\mrm{meta}=M_\mrm{cr}-M_0\sim0.1M_\odot$ \cite{Bombaci:2016xuj}. Combining such a window with the inferred NS birth-mass distribution gives a percent-level fraction over much of the distribution and $f_\mrm{meta}\gtrsim 0.007$ even near its upper tail \cite{You:2024bmk}; thus we adopt $f_\mrm{meta}=10^{-2}$ as a benchmark. The parameter $t_\mrm{meta}$ denotes the time elapsed since a typical star entered the metastable state, not necessarily its full age.

Defining
\begin{align}
C\equiv \frac{\mcal R_\mrm{SGRB}^\mrm{lim}t_\mrm{meta}}
{n_\mrm{NS}f_\mrm{meta}} ~,
\end{align}
the equation $\mcal R_\mrm{tr}=\mcal R_\mrm{SGRB}^\mrm{lim}$ has two positive solutions only for $C<e^{-1}$, i.e.
\begin{align}
\mcal R_\mrm{SGRB}^\mrm{lim}
<\frac{n_\mrm{NS}f_\mrm{meta}}{e\,t_\mrm{meta}} ~.
\label{eq:S_existence}
\end{align}
The boundaries are
\begin{align}
\Gamma_-=-\frac{W_0(-C)}{t_\mrm{meta}} \sep
\Gamma_+=-\frac{W_{-1}(-C)}{t_\mrm{meta}} ~,
\label{eq:S_lambert}
\end{align}
where $W_0$ and $W_{-1}$ are the two real branches of the Lambert $W$ function. The interval $\Gamma_-<\Gamma_\chi<\Gamma_+$ is excluded. The benchmark values in the main text give $C=6.7\times10^{-2}$ and $(\Gamma_-,\Gamma_+)\simeq(7.2\times10^{-11},4.1\times10^{-9})\,\yr^{-1}$.

\section{Effect of neutron star formation}
\label{sec:formation}

The main text neglects the continuous formation of NSs. To estimate its impact, let $\dot n_\mrm{NS}$ be the rate at which NSs enter the old population considered here, and approximate it as constant over $t_\mrm{meta}$. Keeping the present total density $n_\mrm{NS}$ fixed, the transition rate becomes
\begin{align}
\mcal R_\mrm{tr}^{\mrm{form}}
= f_\mrm{meta}\Big[
&(n_\mrm{NS}-\dot n_\mrm{NS}t_\mrm{meta})
\Gamma_\chi e^{-\Gamma_\chi t_\mrm{meta}}
\nonumber\\
&+\dot n_\mrm{NS}
\left(1-e^{-\Gamma_\chi t_\mrm{meta}}\right)\Big] .
\label{eq:S_formation}
\end{align}
The second term accounts for stars added during the interval $t_\mrm{meta}$. A representative low-redshift NS formation rate is $\dot n_\mrm{NS} \sim 10^{-4}\,\Mpc^{-3}\yr^{-1}$, of the order of the core collapse rate and consistent with cosmic remnant estimates \cite{Elbert:2017sbr}. For our benchmark $n_\mrm{NS} = 3\times10^6\,\Mpc^{-3}$ and $t_\mrm{meta} = 1\,\mrm{Gyr}$, only 
\begin{align}
\epsilon\equiv
\frac{\dot n_\mrm{NS}t_\mrm{meta}}{n_\mrm{NS}}
\sim 0.03
\end{align}
of the present population is replenished over this time.

Including Eq.~\eqref{eq:S_formation} shifts the benchmark rate boundaries from $(\Gamma_-,\Gamma_+)\simeq(7.2\times10^{-11}, 4.1\times10^{-9}) \,\yr^{-1}$ to approximately $(7.2 \times 10^{-11}, 5.0 \times 10^{-9})\,\yr^{-1}$. Thus the lower boundary is essentially unchanged, while the upper boundary moves by only $\sim20\%$. In the rapid triggering limit, $\mcal R_\mrm{tr}^{\mrm{form}} \to f_\mrm{meta} \dot n_\mrm{NS} \sim 10^3 \,\mrm{Gpc}^{-3}\yr^{-1}$, still below our benchmark SGRB limit $2\times 10^3 \,\mrm{Gpc}^{-3}\yr^{-1}$. So NS formation gives a smaller correction than the dominant astrophysical uncertainties in $f_\mrm{meta}$ and the transient rate can be neglected at the accuracy of our benchmark analysis. Moreover, formation can only replenish the metastable population and increase the present transition rate, so neglecting it is conservative for the high-$\Gamma_\chi$ constraint.

\section{Microscopic nucleation threshold}
\label{sec:nucleation}

The main text uses the thin-wall potential and barrier in Eqs.~\eqref{bubble_barrier} and \eqref{critical_radius}. To justify the sharp threshold approximation, it is sufficient to note that the collective droplet coordinate has an effective mass \cite{Iida:1998pi,Bombaci:2016xuj}
\begin{align}
M(R)=4\pi\rho_\mrm{H}
\left(1-\frac{n_\mrm{Q}}{n_\mrm{H}}\right)^2R^3 ~,
\end{align}
where $n_\mrm{H,Q}$ are the baryon densities of the two phases and $\rho_\mrm{H}$ is the hadronic mass density. For a quasibound state of energy $E<U_\mrm{max}$, the WKB tunneling probability through the droplet barrier is
\begin{align}
P_\mrm{bub}(E)\simeq
\exp\!\left[-2\int_{R_-}^{R_+}dR\,
\sqrt{2M(R)[U(R)-E]}\right] ~,
\label{eq:S_WKB}
\end{align}
with $U(R_\pm)=E$. The nucleation time scales as $\tau_\mrm{nuc}\sim[\nu N_\mrm{c}P_\mrm{tun}]^{-1}$, where $\nu$ is the oscillation frequency in the inner well and $N_\mrm{c}$ the number of nucleation centers. For metastable matter the exponent is very large when $E$ lies well below the barrier, so prefactor uncertainties are subdominant \cite{Iida:1998pi}.

A localized energy injection shifts the available energy to $E=E_0+ \epsilon _\mrm{bub} E_\mrm{dep}$, where $E_0$ is the ground state energy and $\epsilon_\mrm{bub}$ is the efficiency with which locally deposited energy is transferred to the collective nucleation degree of freedom. Determining this factor requires a nontrivial nonequilibrium QCD calculation, which is beyond the scope of this work. Instead, we assume a maximal-coupling benchmark $\epsilon_\mrm{bub} \sim 1$ following \cite{Herrero:2019esf, AngelesPerez-Garcia:2014cho, Bhutani:2025jfo}. Since the ground-state contribution $E_0$ is small compared with the GeV-scale barriers relevant here, the tunneling suppression is strongly reduced as $E_\mrm{dep}$ approaches $U_\mrm{max}$; above the barrier a supercritical droplet can grow classically. This motivates the step function approximation used in the main text,
$P_\mrm{bub}\simeq\theta(E_\mrm{dep}-U_\mrm{max})$.

This criterion concerns energy deposited within a region of size $R_\mrm{c}$, not the total released energy. The hadronic mean free path quoted in the main text is shorter than $R_\mrm{c}$ for the benchmark parameters. By contrast, applying the standard Landau-Pomeranchuk-Migdal scaling to an electromagnetic shower at nuclear density gives a characteristic length of order $0.6\,\mrm{nm}$ at $E\sim10\,\GeV$ \cite{Graham:2018efk}, approximately five orders of magnitude larger than $R_\mrm{c}$. For this reason we exclude energy that remains in a purely electromagnetic shower; electromagnetic primaries that convert into a sufficiently compact hadronic shower fall within the hadronic case above. Our particle DM limits apply only to the part of the final-state energy that is converted into a sufficiently compact hadronic shower. A channel with a broad deposition spectrum can be treated without the step approximation by retaining the full convolution in Eq.~\eqref{lambda_general}.

\section{Gravitational focusing and particle DM rates}
\label{sec:dm_constraints}

For unbound collisionless DM, Liouville's theorem gives
$f(r,v)=f_\infty(\sqrt{v^2-v_\mrm{esc}^2})$ for $v\ge v_\mrm{esc}(r)$. Writing the normalized asymptotic speed distribution as 
\begin{align}
g_\infty(u) = \frac{4\pi u^2 f_\infty(u)}{n_{\chi,\infty}} \sep
\int_0^\infty g_\infty(u) \,du = 1 ~,
\end{align}
the local density enhancement is
\begin{align}
\frac{n_\chi(r)}{n_{\chi,\infty}}
=\int_0^\infty du\,g_\infty(u)
\frac{\sqrt{u^2+v_\mrm{esc}^2(r)}}{u} ~.
\label{eq:S_focus_general}
\end{align}
For the Maxwellian distribution,
\begin{align}
g_\infty(u)=\frac{4u^2}{\sqrt\pi v_0^3} e^{-u^2/v_0^2} ~,
\end{align}
one obtains \cite{Bromley:2011aa}
\begin{align}
\frac{n_\chi(r)}{n_{\chi,\infty}}
=\frac{2\eta}{\sqrt\pi}+e^{\eta^2}\operatorname{erfc}(\eta)
\xrightarrow{\eta\gg1}\frac{2}{\sqrt\pi}\frac{v_\mrm{esc}}{v_0} \sep
\eta\equiv\frac{v_\mrm{esc}}{v_0} ~,
\label{eq:S_focus_maxwell}
\end{align}
which gives the factor $\simeq9\times10^2$ quoted in the main text for $v_\mrm{esc}=0.6c$ and $v_0=220\,\km/\s$.

Using this focused density and keeping the benchmark volume explicit, the interaction rate normalizations are
\begin{align}
\langle\sigma v\rangle_\mrm{ann} \simeq {}& 5\times10^{-32}\,\cm^3/\s
\left(\frac{m_\chi}{5\,\GeV}\right)^2
\left(\frac{\Gamma_\chi^\mrm{ann}}{4\times 10^{-9}\,\yr^{-1}}\right) \nonumber\\
&\times \left( \frac{0.4\,\GeV/\cm^3}{\rho_{\chi,\infty}} \right)^2 \left(\frac{0.6c}{v_\mrm{esc}}\right)^2
\nonumber\\
&\times \left( \frac{v_0}{220\,\km/\s} \right)^2
\left( \frac{10^6\,\m^3}{V_\mrm{meta}} \right) ~,
\label{eq:S_ann_scaling}
\end{align}
\begin{align}
\tau_\chi\simeq {}& 3 \times 10^{29}\,\s
\left( \frac{10\,\GeV}{m_\chi} \right)
\left( \frac{4\times 10^{-9}\,\yr^{-1}}{\Gamma_\chi^\mrm{dec}} \right) \nonumber\\
&\times\left(\frac{\rho_{\chi,\infty}}{0.4\,\GeV/\cm^3}\right)
\left(\frac{v_\mrm{esc}}{0.6c}\right) \nonumber\\
&\times\left(\frac{220\,\km/\s}{v_0}\right)
\left(\frac{V_\mrm{meta}}{10^6\,\m^3}\right) ~,
\label{eq:S_dec_scaling}
\end{align}
\begin{align}
\sigma_\mrm{inel} \simeq{}& 1 \times10^{-78}\,\cm^2
\left( \frac{m_\chi}{10\,\GeV} \right)
\left( \frac{\Gamma_\chi^\mrm{inel}}{4\times 10^{-9}\,\yr^{-1}} \right) \nonumber\\
&\times \left(\frac{0.4\,\GeV/\cm^3}{\rho_{\chi,\infty}}\right)
\left(\frac{0.6c}{v_\mrm{esc}}\right)^2 \left(\frac{v_0}{220\,\km/\s}\right) \nonumber\\
&\times \left(\frac{0.15\,\mrm{fm}^{-3}}{n_\mrm{b}}\right) \left( \frac{10^6\,\m^3}{V_\mrm{meta}} \right) ~.
\label{eq:S_inel_scaling}
\end{align}
They follow directly from the three event rates displayed in the main text. At the energies of interest, $E_\mrm{dep}\gtrsim U_\mrm{max}\sim10\,\GeV$, Pauli blocking of nucleon final states is negligible for these order of magnitude estimates. Substituting Eq.~\eqref{Gamma_chi_constraint} into Eqs.~\eqref{eq:S_ann_scaling}--\eqref{eq:S_inel_scaling} reproduces Eqs.~\eqref{ann_constraint}--\eqref{inel_constraint}.

\section{Macroscopic dark matter}
\label{sec:macro}

For the optically thick macro limit used in the main text \cite{Jacobs:2014yca, SinghSidhu:2019tbr, Bramante:2026wzh}, the number of nucleon scatters in one critical domain is Poisson distributed with mean $\lambda_\mrm{c} = n_\mrm{b}\sigma_\mrm{macro} R_\mrm{c}$. The exact probability of depositing the required energy in one domain is
\begin{align}
P_\mrm{c}=\sum_{N=N_\mrm{th}}^\infty
\frac{\lambda_\mrm{c}^N}{N!}e^{-\lambda_\mrm{c}} \sep
N_\mrm{th}\simeq\frac{U_\mrm{max}}
{m_N(\gamma_\mrm{esc}^2-1)} ~.
\label{eq:S_macro_Pc}
\end{align}
A path of characteristic length $R_\mrm{meta}$ samples approximately $R_\mrm{meta}/R_\mrm{c}$ domains, so $P_\mrm{th}=1-(1-P_\mrm{c})^{R_\mrm{meta}/R_\mrm{c}}$. In the small-probability and small-$\lambda_\mrm{c}$ limit, this reduces to the expression used in the main text. For the benchmark $N_\mrm{th}=19$, inversion of $\Gamma_\chi^\mrm{macro} = n_{\chi,\mrm{core}}\pi R_\mrm{meta}^2v_\mrm{esc}P_\mrm{th}$ gives
\begin{align}
\sigma_\mrm{macro} \simeq {}& 1.64\times10^{-27}\,\cm^2
\left(\frac{m_\chi}{1\,\g}\right)^\frac{1}{19}
\left(\frac{\Gamma_\chi^\mrm{macro}}{4\times 10^{-9}\,\yr^{-1}}\right)^\frac{1}{19} \nonumber\\
&\times\left(\frac{100\,\m}{R_\mrm{meta}}\right)^\frac{3}{19}
\left(\frac{R_\mrm{c}}{9\,\mrm{fm}}\right)^{-\frac{18}{19}} \left( \frac{0.4\,\GeV/\cm^3}{\rho_{\chi,\infty}} \right)^\frac{1}{19} \nonumber\\
&\times
\left(\frac{v_0}{220\,\km/\s}\right)^\frac{1}{19} \left(\frac{0.6c}{v_\mrm{esc}}\right)^\frac{2}{19}
\left(\frac{0.15\,\mrm{fm}^{-3}}{n_\mrm{b}}\right) ~.
\label{eq:S_macro_scaling}
\end{align}
The weak $1/19$ powers make the inferred cross section insensitive to order-unity changes in the rate normalization. The quoted numerical coefficient uses the leading Poisson tail, $P_\mrm{c}\simeq\lambda_\mrm{c}^{19}/19!$; retaining $e^{-\lambda_\mrm{c}}$ or the exact sum changes the inferred $\sigma_\mrm{macro}$ only weakly because of the nineteenth root.

The deceleration and compactness conditions quoted below Eq.~\eqref{macro_constraint} follow respectively from requiring $\rho_\mrm{b}\sigma_\mrm{macro} R_\mrm{NS}/m_\chi\lesssim1$ and $R_\chi>2Gm_\chi$. The result also assumes an optically thick macro, so that its effective nucleon cross section is geometric; optically thin composite models require replacing $\sigma_\mrm{macro}$ by the appropriate constituent-dependent effective cross section. We do not consider optically thin macros, such as solitons \cite{Zhang:2020bec, Zhang:2021xxa, Visinelli:2021uve, Zhang:2020ntm, Yin:2024xov, Zhang:2024bjo}, in this work.

\end{document}